\documentclass[letterpaper,10pt,conference]{ieeeconf}

\IEEEoverridecommandlockouts
\usepackage{listings}
\usepackage{mathrsfs}
\usepackage{comment}
\usepackage[T1]{fontenc}

\makeatletter
\RequirePackage{amsmath}
\usepackage{mleftright}
\RequirePackage{amssymb}

\RequirePackage{amsthm}
\usepackage{amsthm}
\RequirePackage{bm} 
\usepackage{multirow}
\usepackage{makecell}
\usepackage{booktabs}
\usepackage{array}
\usepackage{url}
\usepackage{graphicx}
\usepackage[caption=false,font=footnotesize]{subfig}
\usepackage{algorithm}
\usepackage{algorithmic}
\usepackage{dsfont}
\usepackage{mathtools}
\usepackage{wrapfig}
\usepackage{lipsum}
\usepackage{epstopdf}
\usepackage{bbm}
\usepackage{relsize}
\usepackage{color}
\usepackage{pifont}
\usepackage{enumerate}
\usepackage{hyperref}
\usepackage{cleveref}

\usepackage{tikz}

\makeatletter
\def\maketag@@@#1{\hbox{\m@th\normalfont\normalsize#1}}
\makeatother

\def\##1\#{\begin{align}#1\end{align}}
\def\$#1\${\begin{align*}#1\end{align*}}

\newtheoremstyle{mytheoremstyle} % name
    {\topsep}                    % Space above
    {\topsep}                    % Space below
    {\normalfont}                   % Body font
    {}                           % Indent amount
    {\bfseries}                   % Theorem head font
    {.}                          % Punctuation after theorem head
    {.5em}                       % Space after theorem head
    {}  % Theorem head spec (can be left empty, meaning ânormalâ)

\theoremstyle{mytheoremstyle}

\ifx\BlackBox\undefined
\newcommand{\BlackBox}{\rule{1.5ex}{1.5ex}}  % end of proof
\fi

\ifx\QED\undefined
\def\QED{~\rule[-1pt]{5pt}{5pt}\par\medskip}
\fi

\ifx\theorem\undefined
\newtheorem{theorem}{Theorem}
\fi
\ifx\example\undefined
\newtheorem{example}[theorem]{Example}
\fi
\ifx\problem\undefined
\newtheorem{problem}[theorem]{Problem}
\fi
\ifx\property\undefined
\newtheorem{property}{Property}
\fi
\ifx\lemma\undefined
\newtheorem{lemma}[theorem]{Lemma}
\fi
\ifx\proposition\undefined
\newtheorem{proposition}[theorem]{Proposition}
\fi
\ifx\remark\undefined
\newtheorem{remark}[theorem]{Remark}
\fi
\ifx\corollary\undefined
\newtheorem{corollary}[theorem]{Corollary}
\fi
\ifx\definition\undefined
\newtheorem{definition}[theorem]{Definition}
\fi
\ifx\conjecture\undefined
\newtheorem{conjecture}[theorem]{Conjecture}
\fi
\ifx\fact\undefined
\newtheorem{fact}[theorem]{Fact}
\fi
\ifx\claim\undefined
\newtheorem{claim}[theorem]{Claim}
\fi
\ifx\assumption\undefined
\newtheorem{assumption}[theorem]{Assumption}
\fi
\numberwithin{equation}{section}
\numberwithin{theorem}{section}

\makeatother

\usepackage{mathtools,amssymb}
\mathtoolsset{showonlyrefs}
\usepackage{graphicx}
\usepackage{pgfplots}
\pgfplotsset{compat=1.18}
\usepgfplotslibrary{fillbetween}
\usetikzlibrary{arrows.meta,positioning}

\DeclareMathOperator{\ERM}{ERM}

\title{\bf Heterogeneous Risk-Sensitive  Evolutionary Learning in  Coordination Games:
Polarization, Welfare Loss, and Intervention  for Efficiency}

\author{Solaleh Mohammadi$^{1}$, Yiman Fong, and Kaiqing Zhang$^{1}$%
\thanks{$^{1}$Solaleh Mohammadi and Kaiqing Zhang are with the University of Maryland, College Park, MD, USA.
Emails: \texttt{solalehm@umd.edu}, \texttt{fangyimin05@gmail.com}, \texttt{kaiqing@umd.edu}. This work was supported by the ARO grant W911NF-24-1-0085, the AFOSR YIP Award
FA9550-25-1-0258, the NSF
CAREER Award 2443704,  and a Cisco Faculty Research Award.}} 

\date{}

\begin{document}
\maketitle
\thispagestyle{empty}
\pagestyle{empty}

\begin{abstract}
We study evolutionary learning in coordination games in a finite population of agents
with heterogeneous risk attitudes, modeled using the entropic risk measure (ERM).
The population consists of two subpopulations, each with a fixed risk attitude.
Under risk-sensitive best response with mutations (BRM), we characterize
the long-run outcomes of the learning process. In particular, we characterize
when a \emph{polarized state}, in which the two subpopulations choose different
actions, is uniquely stochastically stable in terms of agents' risk attitudes
and the subpopulation sizes. Notably,
this polarized state is strictly Pareto dominated by the payoff-dominant equilibrium under \emph{both} expected
payoffs and agents' own \emph{ERM-adjusted payoffs}. Moreover, its welfare, measured by
average expected payoff, is always lower than that of the payoff-dominant
equilibrium and can also be lower than that of the risk-dominant equilibrium.
Thus, heterogeneity
of risk attitudes  in learning alone can lead to an \emph{inefficient} long-run
outcome. To mitigate such inefficiency, we further propose intervention
mechanisms to modify agents' risk attitudes in learning.
We show that a sufficiently large intervention share (the fraction of agents modified)
and sufficiently high intervention intensity (the increase in their risk parameter)
ensure that the payoff-dominant equilibrium is uniquely stochastically stable.
However, greater intervention intensity cannot compensate for an insufficient
intervention share, and a larger intervention share cannot compensate for
insufficient intervention intensity.
\end{abstract}

\section{Introduction}
\label{sec:introduction}

Coordination games, in which agents benefit from matching their actions, provide a classical setting for studying equilibrium selection 
\cite{harsanyi1988general}. Evolutionary game theory, through the framework of stochastic stability, offers a natural way to study which of these equilibrium outcomes emerges. Agents repeatedly revise their actions in response to the population state, while occasional mistakes or experimentation perturb the learning process. Such perturbations can determine which equilibria persist in the long run \cite{foster1990stochastic}.

Learning in coordination games also involves \emph{payoff uncertainty},  because an agent's payoff depends on the action of the opponent she encounters. Agents may differ in how they evaluate the \emph{risk} of this opponent-induced uncertainty. Empirical studies have documented substantial heterogeneity in humans'  risk attitudes \cite{barsky1997preference,choi2007consistency,harrison2007estimating,dohmen2011individual}. Similar heterogeneity can arise in human-robot interaction, where a human may be sensitive to risk while a robot follows a risk-neutral decision rule \cite{kwon2020humans}. This motivates our central question: \emph{how does heterogeneity in agents' risk attitudes shape the long-run outcomes that emerge from evolutionary learning, even when all agents share the same payoff matrix and agree on the preferred outcome?}

Prior work examines the effects of risk attitudes~\cite{nax2019risk,mohammadi2026entropic} and
reference-dependent preferences~\cite{sawa2018reference} on long-run outcomes,
but focuses on homogeneous populations.
Other studies show that heterogeneity in loss aversion can increase common-pool
resource fragility under prospect-theoretic preferences~\cite{hota2016fragility},
while heterogeneity in risk attitudes can change the conditions for strategic
sorting in dynamic conflict games~\cite{van2022fight},  and affect equilibrium contest investment
through the sorting of risk types across
groups~\cite{brookins2021risk}. Related work on coordination and threshold
dynamics links persistent behavioral diversity to several forms of
heterogeneity: opposing rankings of coordinated outcomes can produce
stochastically selected polarization~\cite{neary2012competing}; heterogeneous preferences
can sustain diverse long-run behavior on broad classes of sufficiently large
networks~\cite{neary2017heterogeneity}; heterogeneous best-response thresholds can support the coexistence
of both actions in well-mixed populations~\cite{ramazi2020convergence}; and heterogeneous
observation sample sizes during learning can sustain stable
miscoordination~\cite{arigapudi2025heterogeneous}.

In contrast to these studies, we isolate the effect of heterogeneity in risk attitudes specifically,  by considering a finite population of agents who 
share the same payoff matrix and revision protocol. The underlying
two-player game has two pure strict Nash equilibria. Their population
counterparts, in which all agents choose the same action, are
risk-sensitive equilibria. The \emph{payoff-dominant equilibrium} gives every agent a strictly
higher payoff than the other equilibrium, which is
risk-dominant \cite{harsanyi1988general}. Agents agree on the ranking of these outcomes but
differ in how they evaluate the \emph{risk} arising from opponent-induced
uncertainty. We model these evaluations using the entropic risk measure
(ERM)~\cite{follmer2002convex,howard1972risk}. Each agent has one of two
fixed risk parameters, and agents with the same parameter form a
subpopulation.
We show that heterogeneity in risk attitudes alone can give rise to a
\emph{polarized equilibrium}, in which each subpopulation coordinates on a different action.
Under a risk-sensitive version of \emph{best response with mutations (BRM)} \cite{sandholm2010orders}, we identify when the polarized equilibrium is uniquely stochastically stable in terms of the risk attitudes and relative sizes of the two subpopulations.
Despite the polarized equilibrium's long-run stability, every agent would
be strictly better off at the payoff-dominant equilibrium
under \emph{both} expected payoffs and her own ERM-adjusted payoffs.
Thus, the polarized equilibrium always has lower welfare, measured by average
expected payoff, than the payoff-dominant equilibrium. We also characterize
when its welfare falls below that of the risk-dominant equilibrium.

To address this emerging inefficiency, we study interventions that may \emph{modify}  agents'
risk attitudes in learning to promote the selection of payoff-dominant
equilibrium, motivated by applications of \emph{autonomous agents} that can be controlled/trained by the (human) designers. 
Our prior work~\cite{mohammadi2026entropic} shows that the
payoff-dominant equilibrium is selected in the long run when agents are
sufficiently risk-seeking. Inspired by this result, we consider the
following \emph{intervention} problem: \emph{how can a planner promote selection of the
payoff-dominant equilibrium by choosing a subset of agents and modifying 
their risk attitudes?} To this end, we characterize
the joint roles of \emph{intervention share}, the fraction of the total population whose
risk attitudes are modified, and \emph{intervention intensity}, the
increase in each modified agent's risk parameter. We show that a sufficiently large
intervention share and sufficiently high intervention intensity ensure that the
payoff-dominant equilibrium is uniquely stochastically stable. However, the analysis
also identifies a minimum intervention share that cannot be reduced by increasing
intervention intensity. There is also a minimum intervention intensity required for
efficient selection, even when the intervention share approaches one. We further
show how increasing the intervention share at fixed intervention intensity can
first select a polarized equilibrium and then the payoff-dominant equilibrium.
In addition, we compare modifying  existing agents' risk attitudes with adding new risk-seeking agents and show that, in the large-population limit, modifying existing agents requires the planner to influence fewer agents when both approaches use the same intervention intensity.

\section{Preliminaries}
\label{sec:model}

We study risk-sensitive evolutionary learning in a finite population of $N$
agents with heterogeneous risk attitudes. Agents repeatedly play a two-agent
coordination game with actions $A$ and $B$ against opponents sampled uniformly
from the entire population. They share a payoff matrix but differ in how they
evaluate payoff uncertainty induced by opponents' strategies. Each agent's risk
attitude is fixed over time and does not depend on her current action. At each
time step $t=0,1,2,\ldots$, one uniformly selected agent receives a
\emph{revision opportunity} to update her action, while all others retain theirs.
Revisions follow the risk-sensitive BRM protocol specified in
Section~\ref{subsec:brm-dynamics}.

\subsection{Coordination games in a finite population}

We study a coordination game with an identical payoff matrix for all the agents:
\begin{equation}
\begin{array}{c|cc}
 & A & B\\
\hline
A & a & b\\
B & c & d
\end{array},
\label{eq:coordination-game}
\end{equation}
where $a>c$ and $d>b$. Thus, the action profiles $(A,A)$ and $(B,B)$
are pure strict Nash equilibria of the underlying two-player game.
Among these equilibria, \emph{payoff dominance} means higher payoffs for
both players, whereas \emph{risk dominance} means that the corresponding
action is the strict (risk-neutral)  best response to an opponent equally
likely to choose either action \cite{harsanyi1988general}. Throughout the paper, we impose
Assumption~\ref{ass:payoffs} below to focus on coordination games in which these criteria
favor different equilibria, capturing the non-trivial settings of interest, where there is a tension between higher
payoffs and safer coordination when the opponent's action is
uncertain.

\begin{assumption}[Opposing payoff and risk dominance]
\label{ass:payoffs}
The payoff matrix satisfies:
\renewcommand{\theenumi}{\roman{enumi}}
\renewcommand{\labelenumi}{(\theenumi)}
\begin{enumerate}
\item\label{ass:payoff-dominance}
$a>d$ (payoff dominance of $A$);
\item\label{ass:risk-dominance}
$d-b>a-c$ (risk dominance of $B$).
\end{enumerate}
\end{assumption}

Let $q\in\{0,\ldots,N\}$ denote the number of agents currently choosing action
$A$, and let $x=q/N$ denote the \emph{fraction} of agents in the population choosing $A$. At each revision
opportunity, the selected agent evaluates both actions using
the current fractions $x$ and $1-x$ of agents choosing $A$ and $B$
as the probabilities assigned to the opponent's respective actions. These fractions pool both
subpopulations and count all $N$ agents,
including the selected agent, and are held fixed while she compares the
two actions. We denote the random payoffs from evaluating
$A$ and $B$ by the real-valued random variables $Z_A(x)$ and $Z_B(x)$,
respectively. Specifically, $Z_A(x)$ equals $a$ with probability $x$ and
$b$ with probability $1-x$, while $Z_B(x)$ equals $c$ with probability $x$
and $d$ with probability $1-x$.

Thus, agents face the same payoff matrix and the same distribution of actions
in the population. Heterogeneity enters only through their \emph{evaluation} of these
two random payoffs, as we will elaborate more in \S\ref{sec:risk_intro}.  Finally, to characterize \emph{efficiency} at different values of $x$, we introduce the following notion of welfare.

\begin{definition}[Welfare]
When a fraction $x$ of agents chooses $A$, we define \emph{welfare} as the
average across all $N$ agents of their expected payoffs from their current actions.
In particular, it is given by
\begin{align}
W(x)
&:=x\mathbb E[Z_A(x)]+(1-x)\mathbb E[Z_B(x)]
\nonumber\\
&=ax^2+(b+c)x(1-x)+d(1-x)^2.
\label{eq:welfare}
\end{align}
\end{definition}

\subsection{Payoff evaluation \& Entropic risk attitudes}\label{sec:risk_intro}
When both actions are present in the population, uncertainty about whether
an agent's randomly sampled opponent chooses $A$ or $B$ gives rise to the
random payoffs $Z_A(x)$ and
$Z_B(x)$ defined above. Although agents face the same payoff distributions,
their attitudes toward risk may lead them to evaluate these distributions
differently. We capture these differences through a risk-sensitive payoff
evaluation.
Agent $i$'s risk attitude during learning is represented by the risk parameter
$\beta_i\in\mathbb{R}$. 
For a random payoff $Z$, agent $i$ evaluates $Z$ using the
entropic risk measure~\cite{follmer2002convex,howard1972risk}
\[
\ERM_{\beta_i}(Z):=\frac{1}{\beta_i}
\log\mathbb{E}\!\left[e^{\beta_i Z}\right],\qquad \beta_i\neq0,
\]
with the continuous extension
$\ERM_0(Z)=\mathbb{E}[Z]$.
Under this sign convention, agents with $\beta_i<0$ are risk-averse,
those with $\beta_i=0$ are risk-neutral, and those with $\beta_i>0$
are risk-seeking.

The \emph{ERM-adjusted payoffs} of the two actions are therefore
\begin{align}
U^A_{\beta}(x)
&:=
\frac{1}{\beta}
\log\!\left(
x e^{\beta a}
+(1-x)e^{\beta b}
\right),
\label{eq:risk-value-A}
\\
U^B_{\beta}(x)
&:=
\frac{1}{\beta}
\log\!\left(
x e^{\beta c}
+(1-x)e^{\beta d}
\right).
\label{eq:risk-value-B}
\end{align}
At $\beta=0$, the ERM-adjusted payoffs equal the expected payoffs.
To compare the two actions, we use the ERM-adjusted payoff difference
$g_\beta(x):=U^A_\beta(x)-U^B_\beta(x)$. An agent with risk parameter
$\beta$ strictly prefers $A$ when this difference is positive and $B$
when it is negative; a zero difference means that she is indifferent.

By~\cite[Lemma~1]{mohammadi2026entropic},
each risk parameter $\beta\in\mathbb{R}$ determines a unique best-response 
threshold $\tau(\beta)\in(0,1)$ for the fraction $x$ of agents choosing $A$, in terms of the ERM-adjusted payoffs.
Throughout the paper, threshold refers to a best-response threshold.
An agent with risk
parameter $\beta$ strictly prefers $B$ when $x<\tau(\beta)$, is indifferent
between the two actions when $x=\tau(\beta)$, and strictly prefers $A$ when
$x>\tau(\beta)$.
For $\beta\neq0$,
\[
\tau(\beta)=\frac{e^{\beta d}-e^{\beta b}}
{e^{\beta a}-e^{\beta c}+e^{\beta d}-e^{\beta b}}.
\]
For the risk-neutral case $\beta=0$, we define the threshold as $\tau_0:=\tau(0)=\frac{d-b}{(a-c)+(d-b)}$.
Hence, heterogeneity in risk attitudes induces heterogeneity in the ERM-adjusted best-response
thresholds.

Assumption~\ref{ass:payoffs} implies $c>b$ and $a>d$.
Thus, by~\cite[Lemma~2(II)]{mohammadi2026entropic}, the threshold map
$\tau(\cdot)$ is a continuous bijection from $\mathbb R$ onto $(0,1)$, with
$\frac{\mathrm{d}\tau(\beta)}{\mathrm{d}\beta}<0$, $\lim_{\beta\to-\infty}\tau(\beta)=1$, and
$\lim_{\beta\to+\infty}\tau(\beta)=0$.
Ordering agents or subpopulations by increasing best-response threshold
is therefore equivalent to ordering them by decreasing $\beta$.
Hence, as will be shown later in \S\ref{sec:stochastic-selection}, the analysis of stochastic stability under BRM
will depend on agents' risk parameters
through their best-response thresholds $\tau(\beta)$.

\subsection{Two subpopulations and population state}
\label{subsec:population-states}

Suppose the population consists of two subpopulations of sizes
$n_1=pN$ and $n_2=(1-p)N$, where $p\in(0,1)$ and $n_1,n_2$ are integers.
For $k\in\{1,2\}$, subpopulation $k$ consists of $n_k$ agents who share
the risk parameter $\beta_k$.
Without loss of generality, we label the two subpopulations so that their
best-response thresholds satisfy
\begin{equation}
0<\tau_1<\tau_2<1,
\qquad
\tau_k=\tau(\beta_k).
\label{eq:ordered-two-thresholds}
\end{equation}
Since $\tau(\cdot)$ is strictly decreasing, as recalled in
Section~\ref{sec:risk_intro}, this ordering is equivalent to $\beta_1>\beta_2$.
Hence, agents in subpopulation 1 are more risk-seeking than those in
subpopulation 2. By the best-response rule in the same section, the
payoff-dominant action $A$ is a strict best response for subpopulation 1
when $x>\tau_1$, whereas subpopulation 2 requires $x>\tau_2$.

The \emph{population state} is defined as $s=(i,j)$, where $i\in\{0,\ldots,n_1\}$ and
$j\in\{0,\ldots,n_2\}$ denote the numbers of agents choosing $A$ in
subpopulations 1 and 2, respectively.
The population state space is thus
$\mathcal S=\{0,\ldots,n_1\}\times\{0,\ldots,n_2\}$, and the
fraction of agents choosing $A$ is $x(s)=(i+j)/N$.
With a slight abuse of notation, we write $W(s):=W(x(s))$ to denote the  welfare
at population state $s$.

We call a population state \emph{coordinated} if all agents choose the
same action. The two coordinated states are $C_A=(n_1,n_2)$, where
everyone chooses $A$, and $C_B=(0,0)$, where everyone chooses $B$.
They correspond to uniform adoption of the actions in the pure Nash
equilibria $(A,A)$ and $(B,B)$, respectively. We refer to $C_A$ as
the \emph{payoff-dominant state} and $C_B$ as the \emph{risk-dominant state};
these labels refer to the underlying two-player game and its
risk-neutral payoff comparisons.

We call a population state \emph{polarized} if all agents within each
subpopulation choose the same action, but the two subpopulations
choose different actions. We focus on $P:=(n_1,0)$, in which
subpopulation 1 chooses $A$ and subpopulation 2 chooses $B$, so
$x(P)=p$. The reverse allocation is the polarized state $(0,n_2)$.
The state labels describe action choices; the equilibrium definition
below determines whether these choices are best responses.
To ensure that each agent has a unique best response, we impose the
following assumption throughout the paper.
\begin{assumption}[No ties]
\label{ass:no-ties}
The best-response thresholds satisfy $N\tau_k\notin\mathbb Z$, $k\in\{1,2\}$,
where $\mathbb Z$ denotes the set of integers.
\end{assumption}
Since every feasible fraction of agents choosing $A$ belongs to
$\{0,1/N,\ldots,1\}$, Assumption~\ref{ass:no-ties} ensures that
$x(s)\ne\tau_k$ for every population state $s$ and each subpopulation $k$.
Thus, every agent has a unique best response at every population state.
This is a common, generic assumption: for fixed $N$ and $\tau_k\in(0,1)$,
it excludes only the
finitely many values $\{1/N,\ldots,(N-1)/N\}$ for each threshold.

We define the associated \emph{integer best-response thresholds}, or
\emph{integer thresholds} for short, as $K_k:=\lfloor N\tau_k\rfloor$ for
$k\in\{1,2\}$. These express the best-response rule in terms of the number
$q=i+j$ of agents choosing $A$: under Assumption~\ref{ass:no-ties}, agents
in subpopulation $k$ strictly prefer $A$ when $q\ge K_k+1$ and $B$ when
$q\le K_k$.
For an agent in subpopulation $k$, we denote her unique
\emph{risk-sensitive best response} at a feasible fraction $x$ by
$BR_{\beta_k}(x)$, with $BR_{\beta_k}(x)=A$ if $x>\tau_k$ and
$BR_{\beta_k}(x)=B$ if $x<\tau_k$.

\begin{definition}[Risk-sensitive equilibrium]
\label{def:risk-sensitive-equilibrium}
A population state $s$ is a \emph{risk-sensitive equilibrium} if every
agent in each subpopulation $k$ chooses $BR_{\beta_k}(x(s))$.
Thus, each agent's current action maximizes her ERM-adjusted payoff,
with the population fraction $x(s)$ held fixed when comparing actions.
\end{definition}

In the analysis of our population model, \emph{equilibrium} refers to
a risk-sensitive equilibrium unless otherwise specified. When $P$
satisfies Definition~\ref{def:risk-sensitive-equilibrium}, we call it
a \emph{polarized equilibrium}. Proposition~\ref{prop:two-type-recurrent}
shows that $C_A$ and $C_B$ are equilibria for all risk parameters and
characterizes when $P$ is also an equilibrium.

\subsection{Motivating example: Common payoffs but  heterogeneous risk attitudes}
\label{subsec:motivating-example}

\begin{figure}[t]
% Keep the top axis border inside the conference text area.
\vspace*{5pt}
\centering
% Requires pgfplots (compat=1.18) and the TikZ arrows.meta library.
% Best-response regions in x, not stochastic-selection regions in p.
\begingroup
\definecolor{riskSeekingCurve}{HTML}{0077CC}
\definecolor{riskAverseCurve}{HTML}{F07818}
\definecolor{preferBRegion}{HTML}{C94C4C}
\definecolor{opposingRegion}{HTML}{EBCB56}
\definecolor{preferARegion}{HTML}{75B9EB}
\pgfmathsetmacro{\riskTauOne}{(exp(-4)+exp(-6))/(1+exp(-4)+exp(-6))}
\pgfmathsetmacro{\riskTauTwo}{(exp(1)+exp(1.5))/(1+exp(1)+exp(1.5))}
\def\riskShare{0.30}
\begin{tikzpicture}[
    declare function={
        riskgap(\x,\b)=ln((\x*exp(4*\b)+(1-\x))
            /(\x*exp(3*\b)+(1-\x)*exp(2*\b)))/\b;
    },
    risk label/.style={font=\scriptsize, inner sep=1.5pt, align=center},
    risk arrow/.style={-{Latex[length=1.3mm]}, thin}
]
\begin{axis}[
    width=\columnwidth,
    height=0.60\columnwidth,
    xmin=0, xmax=1,
    ymin=-2.05, ymax=1.22,
    axis on top,
    tick label style={font=\scriptsize},
    label style={font=\footnotesize},
    xlabel={Population share $x$ choosing $A$},
    ylabel={$g_\beta(x)=U^A_\beta(x)-U^B_\beta(x)$},
    xtick={0,0.2,0.4,0.6,0.8,1},
    ytick={-2,-1,0,1},
    legend style={
        at={(0.97,0.04)}, anchor=south east,
        font=\scriptsize, draw=black!20, fill=white,
        cells={anchor=west}, inner sep=2pt
    },
]

% Use the same thresholds for the shading, guide lines, and zero markers.
\path[fill=preferBRegion!18]
    (axis cs:0,-2.05) rectangle (axis cs:\riskTauOne,1.22);
\path[fill=opposingRegion!18]
    (axis cs:\riskTauOne,-2.05) rectangle (axis cs:\riskTauTwo,1.22);
\path[fill=preferARegion!22]
    (axis cs:\riskTauTwo,-2.05) rectangle (axis cs:1,1.22);
\draw[black!30, densely dotted]
    (axis cs:\riskTauOne,-2.05) -- (axis cs:\riskTauOne,1.22)
    (axis cs:\riskTauTwo,-2.05) -- (axis cs:\riskTauTwo,1.22);
\draw[black!65, thin] (axis cs:0,0) -- (axis cs:1,0);
\draw[black!65, dashed]
    (axis cs:\riskShare,-2.05) -- (axis cs:\riskShare,1.22);

% Quadratic sampling resolves the steep risk-seeking curve near x=0.
\addplot[riskSeekingCurve, thick, domain=0:1, samples=251]
    ({x^2},{riskgap(x^2,2)});
\addlegendentry{$\beta_1=2$}
\addplot[riskAverseCurve, thick, domain=0:1, samples=201]
    {riskgap(x,-0.5)};
\addlegendentry{$\beta_2=-0.5$}

\addplot[only marks, mark=*, mark size=1.7pt,
    mark options={fill=black, draw=white}, forget plot]
    coordinates {(\riskTauOne,0) (\riskTauTwo,0)};
\addplot[only marks, mark=*, mark size=2pt,
    mark options={fill=riskSeekingCurve, draw=white}, forget plot]
    coordinates {(\riskShare,{riskgap(\riskShare,2)})};
\addplot[only marks, mark=*, mark size=2pt,
    mark options={fill=riskAverseCurve, draw=white}, forget plot]
    coordinates {(\riskShare,{riskgap(\riskShare,-0.5)})};

\node[risk label, anchor=south west] at (axis cs:0.10,0.18)
    {$\tau_1\approx0.0204$};
\draw[risk arrow] (axis cs:0.10,0.18) -- (axis cs:\riskTauOne,0.025);
\node[risk label, anchor=south east] at (axis cs:0.85,0.30)
    {$\tau_2\approx0.8780$};
\draw[risk arrow] (axis cs:0.84,0.30) -- (axis cs:\riskTauTwo,0.025);

% The narrow red region needs a callout to keep its label legible.
\node[risk label, text=preferBRegion!80!black, anchor=west]
    at (axis cs:0.055,-0.72) {Both prefer $B$};
\draw[risk arrow, draw=preferBRegion!80!black]
    (axis cs:0.075,-0.85) -- (axis cs:{\riskTauOne/2},-1.10);
\node[risk label, text=black!80] at (axis cs:0.57,-0.48)
    {Opposing best responses\\$\tau_1<x<\tau_2$};
\node[risk label, text=riskSeekingCurve!80!black] at (axis cs:0.94,-0.65)
    {Both\\prefer\\$A$};

\node[risk label, anchor=north, fill=opposingRegion!18]
    at (axis cs:\riskShare,1.20) {$p=0.30$};
\node[risk label, text=riskSeekingCurve, anchor=north west]
    at (axis cs:0.34,0.80) {$g_2(p)>0$};
\node[risk label, text=riskAverseCurve!85!black, anchor=north west]
    at (axis cs:0.34,-1.60) {$g_{-0.5}(p)<0$};
\end{axis}
\end{tikzpicture}
\endgroup
\caption{ERM-adjusted payoff differences and best-response regions.}
\label{fig:different-risk-attitudes}
\end{figure}
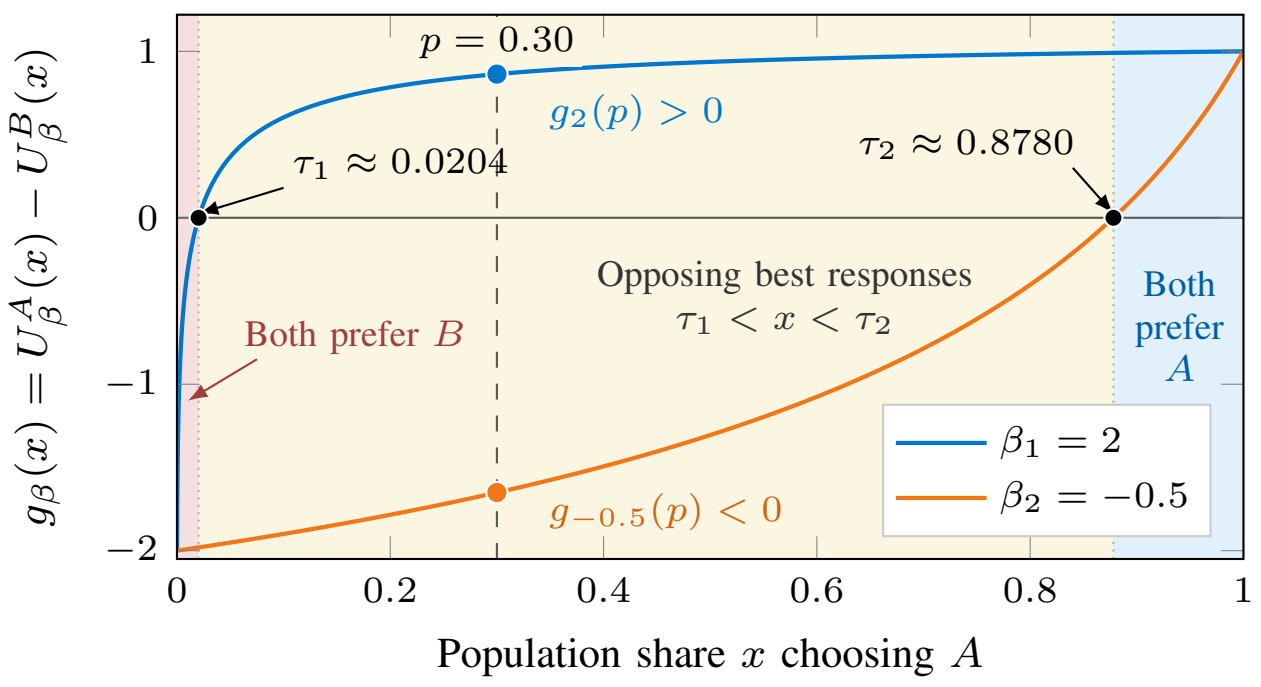

Consider the coordination game with payoff matrix 
\begin{equation}
\begin{array}{c|cc}
 & A & B\\
\hline
A & 4 & 0\\
B & 3 & 2
\end{array}.
\label{eq:motivating-game}
\end{equation}
In this example, risk-seeking agents account for a fraction $p=0.30$ of the
population and have risk parameter $\beta_1=2$ and thus best-response threshold
$\tau_1=\tau(2)\approx0.0204$. The remaining fraction $1-p=0.70$
consists of risk-averse agents with $\beta_2=-0.5$ and threshold
$\tau_2=\tau(-0.5)\approx0.8780$.
Consider the polarized state $P=(n_1,0)$, in which all risk-seeking
agents choose $A$ and all risk-averse agents choose $B$.
The fraction of agents choosing $A$ is then
$x=p=0.30$. Since $\tau_1<p<\tau_2$, $A$ is a strict best response for the
risk-seeking agents, while $B$ is a strict best response for the risk-averse
agents. Fig.~\ref{fig:different-risk-attitudes} illustrates these opposing
best responses.
Thus, $P$ is a risk-sensitive equilibrium by
Definition~\ref{def:risk-sensitive-equilibrium}, illustrating the
polarized equilibrium characterized in Proposition~\ref{prop:two-type-recurrent}.

By Eq.~\eqref{eq:welfare}, welfare at the polarized equilibrium is
$W(0.30)=4(0.30)^2+3(0.30)(0.70)+2(0.70)^2=1.97$.
Since $W(0)=2$ and $W(1)=4$, the polarized equilibrium has lower welfare
than either $C_B$ or $C_A$.
This polarization arises solely from heterogeneous risk attitudes.
Although every agent would prefer the outcome in which all agents choose
$A$, her current action in the polarized equilibrium is already a strict
risk-sensitive best response. Once the population reaches this equilibrium,
subsequent best-response revisions therefore leave it there. In the next
section, we study whether this inefficient equilibrium is
stochastically stable under BRM learning dynamics, which allow occasional
mistakes in agents' action choices.

\section{Learning Dynamics with Heterogeneous Risk Attitudes}
\label{sec:learning-dynamics}

We now specify risk-sensitive BRM dynamics for the population model
introduced in Section~\ref{subsec:population-states}. Agents revise their
actions while their risk parameters remain fixed, and we study which
risk-sensitive equilibria persist as mistakes become rare.

\subsection{Risk-sensitive BRM learning dynamics}
\label{subsec:brm-dynamics}

Our protocol applies the BRM update rule~\cite{sandholm2010orders}
to agents' ERM-adjusted best responses. We refer to this risk-sensitive
protocol as BRM below and analyze its long-run behavior using the
stochastic stability framework of~\cite{kandori1993learning,young1993evolution}.

Consider a mutation probability $\varepsilon\in(0,1/2)$, which  represents the probability of  rare mistakes
or experimentation by the agents during learning, and is the same for every agent and population state.
At each time step, one agent is selected uniformly at random from all $N$
agents to revise her action. Each agent is selected with probability $1/N$,
so the selected agent belongs to subpopulation $k$ with probability $n_k/N$
(namely, $p$ for subpopulation 1 and $1-p$ for subpopulation 2).
She chooses her risk-sensitive best response with probability
$1-\varepsilon$ and the other action with probability
$\varepsilon$; all other agents retain their actions.
For a selected agent in subpopulation $k$, the probability of choosing
$A$ at population state $s$ is
\begin{equation}
b_k^\varepsilon(s)
:=
\begin{cases}
1-\varepsilon, & x(s)>\tau_k,\\
\varepsilon, & x(s)<\tau_k.
\end{cases}
\label{eq:perturbed-BR}
\end{equation}
The probability of choosing $B$ is thus $1-b_k^\varepsilon(s)$.

Let $s(t)=(i(t),j(t))\in\mathcal S$ denote the population state at time step 
$t$. Note that the numbers of agents using each action in each subpopulation and
their revision probabilities depend only on $s(t)$. Thus, $\{s(t)\}_{t\ge0}$
is a finite-state Markov chain defined on $\mathcal S$.

Let $M^\varepsilon$ denote the \emph{transition matrix} of the Markov chain $\{s(t)\}_{t\ge0}$
associated with the BRM dynamics with mutation probability
$\varepsilon$. 
Its entry $M^\varepsilon(s,s')$ denotes the probability that one BRM revision
takes the population from $s$ to $s'$. A transition $s\to s'$ is
\emph{feasible} if $M^\varepsilon(s,s')>0$ for every
$\varepsilon\in(0,1/2)$. Write $s_1=i$, $s_2=j$, and let
$e_k$ be the $k$th coordinate vector. For $k\in\{1,2\}$, there are $n_k-s_k$
agents choosing $B$ and $s_k$ agents choosing $A$ in subpopulation $k$.
Since the revising agent is drawn uniformly from all $N$ agents, the
probabilities of selecting an agent from these two groups are
$(n_k-s_k)/N$ and $s_k/N$, respectively. Multiplying by the conditional
probability of switching actions gives the transition probabilities to
feasible neighboring states:
\begin{align*}
M^\varepsilon(s,s+e_k)
&=\frac{n_k-s_k}{N}\,b_k^\varepsilon(s),\\
M^\varepsilon(s,s-e_k)
&=\frac{s_k}{N}\bigl(1-b_k^\varepsilon(s)\bigr).
\end{align*}
All other transitions to distinct states have probability zero, and
\[
M^\varepsilon(s,s):=1-\sum_{s'\ne s}M^\varepsilon(s,s').
\]
The \emph{zero-noise transition matrix} $M^0$ is defined by the same rule
with $\varepsilon=0$: the selected agent always chooses her
risk-sensitive best response.

For every $\varepsilon\in(0,1/2)$, each feasible single-coordinate
increase or decrease has positive probability. Such moves connect all
population states in $\mathcal S$, making the chain irreducible. Every
selected agent also has positive probability of retaining her current
action, giving a positive self-transition probability at every population
state and hence aperiodicity of the Markov chain. 
For $\varepsilon\in(0,1/2)$, let $\pi^\varepsilon$ denote the unique
stationary distribution of $M^\varepsilon$ on $\mathcal S$, whose
existence and uniqueness follow from finiteness and irreducibility.
Aperiodicity further ensures that, for each fixed $\varepsilon\in(0,1/2)$
and every initial distribution $\mu$ on $\mathcal S$,
$\mu(M^\varepsilon)^t\to\pi^\varepsilon$ as $t\to\infty$.
See~\cite{stroock2005introduction} for these standard Markov chain results.

\subsection{Equilibria and stochastic stability}

Define the \emph{zero-noise graph} $\mathcal G^0$ as the directed graph
with vertex set $\mathcal S$ and an edge $s\to s'$ whenever $s\ne s'$ and
$M^0(s,s')>0$. Its edges represent single best-response action
changes. A state is \emph{absorbing} if it has no outgoing edge.
A \emph{communicating class} is a maximal set of mutually reachable
states; it is \emph{closed} if no edge leaves it. These closed classes
are the recurrent classes of the Markov chain on $\mathcal S$ with
transition matrix $M^0$~\cite{young1993evolution}.

Under Assumption~\ref{ass:no-ties}, a population state satisfies
Definition~\ref{def:risk-sensitive-equilibrium} exactly when it is
absorbing for the Markov chain with transition matrix $M^0$, that is, $M^0(s,s)=1$.
At such a state, every selected agent retains her current action
because it is her unique ERM-adjusted best response.

\begin{definition}[Stochastic stability]
Following~\cite{young1993evolution}, a population state $s$ is \emph{stochastically stable} if
$\lim_{\varepsilon\to0}\pi^\varepsilon(s)>0$, and \emph{uniquely stochastically stable}
if $\lim_{\varepsilon\to0}\pi^\varepsilon(s)=1$.
An equilibrium that is stochastically stable is called a
\emph{stochastically stable equilibrium} (SSE).
\end{definition}

The stochastic-stability question therefore asks which equilibrium
retains positive stationary probability when mistakes become
arbitrarily rare.

\begin{proposition}[Risk-sensitive equilibria under zero noise]
\label{prop:two-type-recurrent}
Suppose Eq.~\eqref{eq:ordered-two-thresholds} and Assumption~\ref{ass:no-ties}
hold. The states $C_B$ and $C_A$ are risk-sensitive equilibria.
The polarized state $P$ is a risk-sensitive equilibrium if and only if
\begin{equation}
K_1+1\le n_1\le K_2,
\label{eq:P-recurrent-finite}
\end{equation}
or, equivalently, $\tau_1<p<\tau_2$.
There are no other risk-sensitive equilibria. The recurrent classes
of the Markov chain with transition matrix $M^0$ are exactly $\{C_B\}$ and $\{C_A\}$,
together with $\{P\}$ when Eq.~\eqref{eq:P-recurrent-finite} holds.
The chain reaches one of these absorbing states almost surely from
every initial state.
\end{proposition}

\begin{proof}
At $C_B$, we have $x=0$, so both actions yield deterministic payoffs
against the current population. Since the ERM of a deterministic payoff
equals that payoff for every $\beta$, we have
$U^B_\beta(0)=d>b=U^A_\beta(0)$.
Similarly, at $C_A$, we have $x=1$ and
$U^A_\beta(1)=a>c=U^B_\beta(1)$.
Thus, every agent's current action is her strict ERM-adjusted best
response at both states, making $C_B$ and $C_A$ risk-sensitive
equilibria for all risk parameters.
At $P$, the fraction of agents choosing $A$ is $x=p$. Every agent strictly prefers her
current action exactly when
$U^A_{\beta_1}(p)>U^B_{\beta_1}(p)$ and
$U^B_{\beta_2}(p)>U^A_{\beta_2}(p)$.
By the best-response threshold rule in Section~\ref{sec:risk_intro},
these inequalities hold if and only if $\tau_1<p<\tau_2$.
Under this condition, every agent chooses her unique ERM-adjusted
best response, so $P$ is a risk-sensitive equilibrium and $M^0(P,P)=1$.
Since $n_1=pN$ and $N\tau_1,N\tau_2\notin\mathbb Z$ by
Assumption~\ref{ass:no-ties},
the condition is equivalent to Eq.~\eqref{eq:P-recurrent-finite}.

To exclude other equilibria, observe that agents within each subpopulation
have the same unique best response under Assumption~\ref{ass:no-ties}.
Thus, at equilibrium each subpopulation must use a single action.
Besides $C_B$, $C_A$, and $P$, the only candidate is the reversed
polarized state $(0,n_2)$. At this state, subpopulation 1 chooses $B$,
which requires $x<\tau_1$ to be a best response, while subpopulation 2
chooses $A$, which requires $x>\tau_2$. These conditions cannot both
hold because $\tau_1<\tau_2$.

It remains to exclude recurrent classes with more than one state.
Let $q:=i+j$ and define
$\Psi(i,j):=\frac{q(q-1)}{2}-K_1 i-K_2 j$.
By Assumption~\ref{ass:no-ties}, an agent in subpopulation $k$ strictly prefers $A$
when $q\ge K_k+1$ and $B$ when $q\le K_k$. If she switches from
$B$ to $A$, the change in $\Psi$ is
$\Delta\Psi=q-K_k\ge1$. If she switches from $A$ to $B$, it is
$\Delta\Psi=K_k-q+1\ge1$.
Here $q$ is the aggregate count before the revision. Thus, $\Psi$
strictly increases along every edge of $\mathcal G^0$, so the graph
has no directed cycles and no two distinct states can communicate.
Its closed communicating classes are therefore absorbing singletons,
which are precisely the equilibria identified above. Finally, a finite
Markov chain reaches a recurrent class almost surely~\cite{stroock2005introduction}, proving the
claimed absorption from every initial state.
\end{proof}

\subsection{Minimum resistances between equilibria}
\label{subsec:minimum-resistances}

Following~\cite{young1993evolution}, we call $M^\varepsilon$ a
\emph{regular perturbation} of $M^0$ if it is irreducible and
aperiodic for sufficiently small positive $\varepsilon$, converges
entrywise to $M^0$ as $\varepsilon\to0$, and for every feasible
transition $s\to s'$ there exists a nonnegative real number
$\rho(s,s')$ satisfying
\begin{equation}
0<\lim_{\varepsilon\to 0}
\frac{M^\varepsilon(s,s')}{\varepsilon^{\rho(s,s')}}<\infty.
\label{eq:transition-resistance}
\end{equation}
We call $\rho(s,s')$ the \emph{resistance}  of the transition $s\to s'$
and set $\rho(s,s'):=+\infty$ for impossible transitions.

\begin{proposition}[Mistakes as resistances]
\label{prop:mistake-resistance}
Under Assumption~\ref{ass:no-ties}, the BRM
protocol in Eq.~\eqref{eq:perturbed-BR} defines a regular perturbation of $M^0$.
For every feasible transition between distinct states, $\rho(s,s')=0$
if the revision is a best response and $\rho(s,s')=1$ if it is a mistake.
\end{proposition}
\begin{proof}
A transition between distinct states changes one agent's action within
one subpopulation.
If $m\ge1$ agents can make this transition, the probability of selecting
one of them is $m/N$, since selection is uniform over the entire population.
The transition probability is therefore
$(m/N)(1-\varepsilon)$ for a best response and $(m/N)\varepsilon$
for a mistake. For a best response,
$\lim_{\varepsilon\to0}M^\varepsilon(s,s')=m/N>0$, so
Eq.~\eqref{eq:transition-resistance} holds with $\rho(s,s')=0$.
For a mistake, dividing the transition probability by $\varepsilon$
gives $M^\varepsilon(s,s')/\varepsilon=m/N>0$, so the same condition
holds with $\rho(s,s')=1$.
Self-transition probabilities
are sums of such terms and have well-defined leading powers.
Every individual action change has positive probability when
$0<\varepsilon<1/2$, so the state space communicates. Moreover,
$M^\varepsilon(C_B,C_B)=1-\varepsilon>0$, giving aperiodicity.
Finally, the transition probabilities converge to those of $M^0$.
These facts establish that $M^\varepsilon$ is a regular perturbation of $M^0$.
\end{proof}

Consequently, for distinct states $s,s'$, an edge $s\to s'$ belongs
to $\mathcal G^0$ exactly when $\rho(s,s')=0$. Paths in
$\mathcal G^0$ therefore consist entirely of zero-resistance revisions.
For states $u,v\in\mathcal S$, a \emph{path} from $u$ to $v$ is a finite
sequence of states $\omega=(s_0,\ldots,s_L)$ that starts at $s_0=u$ and
ends at $s_L=v$. Each step $s_\ell\to s_{\ell+1}$ must be a feasible
transition, for $\ell=0,\ldots,L-1$, where $L$ is the number of transitions
along the path. We denote such a path by $\omega:u\leadsto v$ and define
its \emph{path resistance} $R(\omega)$ and the \emph{minimum resistance}
$r(u,v)$ from $u$ to $v$ by
\[
\begin{aligned}
R(\omega):=\sum_{\ell=0}^{L-1}\rho(s_\ell,s_{\ell+1}),~~
r(u,v):=\min_{\omega:u\leadsto v}R(\omega).
\end{aligned}
\]
By Proposition~\ref{prop:mistake-resistance}, $r(u,v)$ is the fewest
mistakes needed to reach $v$ from $u$, attained on a path without repeated
states.

\begin{proposition}[Minimum resistances]
\label{prop:two-type-minimum-resistances}
Suppose Eq.~\eqref{eq:ordered-two-thresholds},
Assumption~\ref{ass:no-ties}, and $\tau_1<p<\tau_2$ hold. The minimum
resistances between $P$ and each of $C_B$ and $C_A$ are
\begin{align}
r(C_B,P)&=K_1+1,\label{eq:r-BP}\\
r(P,C_B)&=n_1-K_1,\label{eq:r-PB}\\
r(P,C_A)&=K_2+1-n_1,\label{eq:r-PA}\\
r(C_A,P)&=N-K_2.\label{eq:r-AP}
\end{align}
\par\smallskip\noindent
The minimum resistances between $C_B$ and $C_A$ are
\begin{align}
r(C_B,C_A)
&=
\max\left\{
K_1+1,\,
K_2+1-n_1
\right\},
\label{eq:r-BA-direct}
\\
r(C_A,C_B)
&=
\max\left\{
N-K_2,\,
n_1-K_1
\right\}.
\label{eq:r-AB-direct}
\end{align}
\end{proposition}

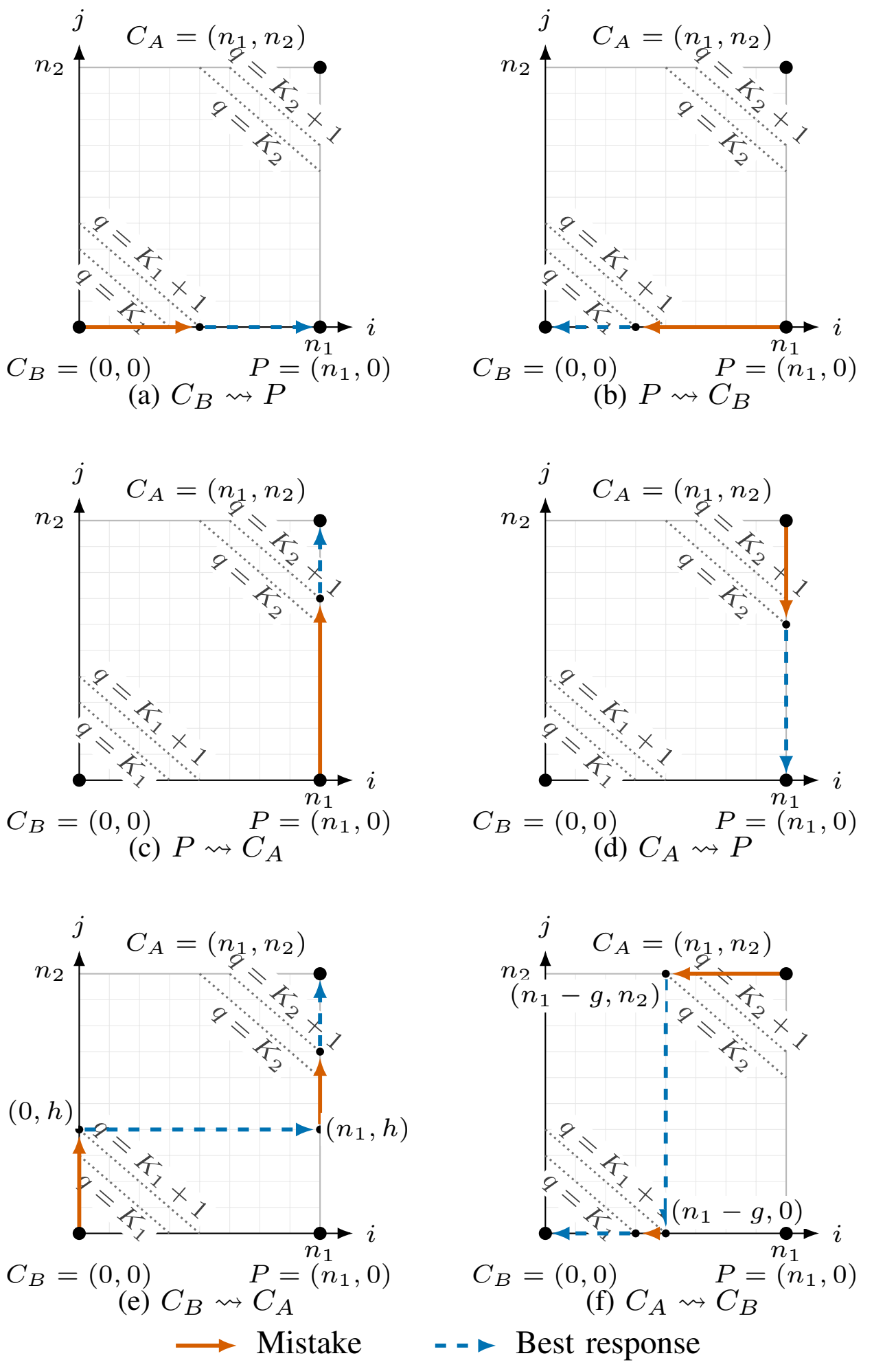
\begin{figure}[t]
\centering
\begingroup
\definecolor{resistanceMistake}{HTML}{D55E00}
\definecolor{resistanceBest}{HTML}{0072B2}
\tikzset{
    resistance diagram/.style={x=0.29cm,y=0.25cm,font=\scriptsize},
    resistance mistake/.style={
        draw=resistanceMistake,line width=1.1pt,
        -{Latex[length=1.8mm]},shorten >=1.5pt,shorten <=1.5pt
    },
    resistance best/.style={
        draw=resistanceBest,line width=1.1pt,dashed,
        -{Latex[length=1.8mm]},shorten >=1.5pt,shorten <=1.5pt
    },
    resistance threshold/.style={
        draw=black!55,text=black!75,densely dotted,line width=0.6pt,
        font=\fontsize{7}{8}\selectfont
    },
    resistance label/.style={fill=white,inner sep=1pt}
}
% The same integer state space is used in all six panels.
\newcommand{\resistanceStateAxes}{%
    \path[use as bounding box] (-2.5,-1.5) rectangle (11,12.3);
    \draw[step=1,gray!18,very thin] (0,0) grid (8,10);
    \draw[gray!55,thin] (0,10)--(8,10)--(8,0);
    \draw[-{Latex[length=1.5mm]},thin] (0,0)--(9.1,0) node[right] {$i$};
    \draw[-{Latex[length=1.5mm]},thin] (0,0)--(0,10.9) node[above] {$j$};
    \draw[resistance threshold] (0,3)--(3,0)
        node[pos=0.5,sloped,below,resistance label] {$q=K_1$};
    \draw[resistance threshold] (0,4)--(4,0)
        node[pos=0.5,sloped,above,resistance label] {$q=K_1+1$};
    \draw[resistance threshold] (4,10)--(8,6)
        node[pos=0.5,sloped,below,resistance label] {$q=K_2$};
    \draw[resistance threshold] (5,10)--(8,7)
        node[pos=0.5,sloped,above,resistance label] {$q=K_2+1$};
    \node[left] at (0,10) {$n_2$};
    \node[below] at (8,0) {$n_1$};
}
\newcommand{\resistanceStateLabels}{%
    \fill (0,0) circle (1.8pt);
    \fill (8,0) circle (1.8pt);
    \fill (8,10) circle (1.8pt);
    \node[below] at (0,-0.65) {$C_B=(0,0)$};
    \node[below] at (8,-0.65) {$P=(n_1,0)$};
    \node[above left] at (8,10.15) {$C_A=(n_1,n_2)$};
}
\subfloat[$C_B\leadsto P$\label{fig:state-path-BP}]{%
\begin{minipage}[t]{0.48\linewidth}
\centering
\begin{tikzpicture}[resistance diagram]
\resistanceStateAxes
\draw[resistance mistake] (0,0)--(4,0);
\draw[resistance best] (4,0)--(8,0);
\fill (4,0) circle (1.1pt);
\resistanceStateLabels
\end{tikzpicture}
\end{minipage}}\hfill
\subfloat[$P\leadsto C_B$\label{fig:state-path-PB}]{%
\begin{minipage}[t]{0.48\linewidth}
\centering
\begin{tikzpicture}[resistance diagram]
\resistanceStateAxes
\draw[resistance mistake] (8,0)--(3,0);
\draw[resistance best] (3,0)--(0,0);
\fill (3,0) circle (1.1pt);
\resistanceStateLabels
\end{tikzpicture}
\end{minipage}}
\par\smallskip
\subfloat[$P\leadsto C_A$\label{fig:state-path-PA}]{%
\begin{minipage}[t]{0.48\linewidth}
\centering
\begin{tikzpicture}[resistance diagram]
\resistanceStateAxes
\draw[resistance mistake] (8,0)--(8,7);
\draw[resistance best] (8,7)--(8,10);
\fill (8,7) circle (1.1pt);
\resistanceStateLabels
\end{tikzpicture}
\end{minipage}}\hfill
\subfloat[$C_A\leadsto P$\label{fig:state-path-AP}]{%
\begin{minipage}[t]{0.48\linewidth}
\centering
\begin{tikzpicture}[resistance diagram]
\resistanceStateAxes
\draw[resistance mistake] (8,10)--(8,6);
\draw[resistance best] (8,6)--(8,0);
\fill (8,6) circle (1.1pt);
\resistanceStateLabels
\end{tikzpicture}
\end{minipage}}
\par\smallskip
\subfloat[$C_B\leadsto C_A$\label{fig:state-path-to-A}]{%
\begin{minipage}[t]{0.48\linewidth}
\centering
\begin{tikzpicture}[resistance diagram]
\resistanceStateAxes
\draw[resistance mistake] (0,0)--(0,4);
\draw[resistance best] (0,4)--(8,4);
\draw[resistance mistake] (8,4)--(8,7);
\draw[resistance best] (8,7)--(8,10);
\foreach \x/\y in {0/4,8/4,8/7}
    \fill (\x,\y) circle (1.1pt);
\node[above left,resistance label] at (0,4) {$(0,h)$};
\node[right,resistance label] at (8,4) {$(n_1,h)$};
\resistanceStateLabels
\end{tikzpicture}
\end{minipage}}\hfill
\subfloat[$C_A\leadsto C_B$\label{fig:state-path-to-B}]{%
\begin{minipage}[t]{0.48\linewidth}
\centering
\begin{tikzpicture}[resistance diagram]
\resistanceStateAxes
\draw[resistance mistake] (8,10)--(4,10);
\draw[resistance best] (4,10)--(4,0);
\draw[resistance mistake] (4,0)--(3,0);
\draw[resistance best] (3,0)--(0,0);
\foreach \x/\y in {4/10,4/0,3/0}
    \fill (\x,\y) circle (1.1pt);
\node[below left,resistance label] at (4,9.8) {$(n_1-g,n_2)$};
\node[above right,resistance label] at (4,0.15) {$(n_1-g,0)$};
\resistanceStateLabels
\end{tikzpicture}
\end{minipage}}
\par\smallskip
{\small
\tikz[baseline=-0.5ex]{\draw[resistance mistake] (0,0)--(0.7,0);}
Mistake\qquad
\tikz[baseline=-0.5ex]{\draw[resistance best] (0,0)--(0.7,0);}
Best response
}
\endgroup
\caption{Minimum-resistance paths between $C_B$, $P$, and $C_A$.}
\label{fig:state-space-paths}
\end{figure}

\begin{proof}
Recall that $q=i+j$ and $K_1+1\le n_1\le K_2$. By
Proposition~\ref{prop:mistake-resistance}, a path's resistance is its
number of mistakes. A given agent is selected and switches actions with
probability at least $\varepsilon/N>0$, so the finite sequences below
are possible under BRM. Fig.~\ref{fig:state-space-paths} illustrates
the state space: horizontal moves change actions in subpopulation 1,
and vertical moves change actions in subpopulation 2.

For $C_B\leadsto P$, consider a path on which only subpopulation 1
switches to $A$, along the bottom edge $j=0$ in
Fig.~\ref{fig:state-path-BP}.
The first $K_1+1$ switches are mistakes;
best responses complete the path. This count is minimal because every
path must first reach a state with $q=K_1+1$ from zero, and all switches
to $A$ up to that visit are mistakes. For $P\leadsto C_B$
(Fig.~\ref{fig:state-path-PB}), only
subpopulation 1 switches to $B$: $n_1-K_1$ mistakes lead to $(K_1,0)$,
and best responses lead to $C_B$. Any path must first reach a state
with $q=K_1$, where $i\le K_1$. It therefore requires at least
$n_1-K_1$ switches to $B$ in subpopulation 1 while $q>K_1$, all
mistakes. This proves Eqs.~\eqref{eq:r-BP} and~\eqref{eq:r-PB}.

For $C_A\leadsto P$
(Fig.~\ref{fig:state-path-AP}),
$N-K_2$ mistaken switches to $B$ in
subpopulation 2 follow the right edge $i=n_1$ to $(n_1,K_2-n_1)$;
best responses in that
subpopulation then lead to $P$. Every path must first reach $q=K_2$
from $N$ while both subpopulations prefer $A$, requiring at least
$N-K_2$ mistakes. This proves Eq.~\eqref{eq:r-AP}.
For $P\leadsto C_A$
(Fig.~\ref{fig:state-path-PA}),
$K_2+1-n_1$ mistaken switches to $A$ in
subpopulation 2 raise $q$ to $K_2+1$, after which best responses
lead to $C_A$. Every path's first state with $q=K_2+1$ has
$j\ge K_2+1-n_1$. Since $j$ starts at zero, this requires at least
that many switches to $A$ in subpopulation 2 while $q\le K_2$, all
mistakes. This proves Eq.~\eqref{eq:r-PA}.

For $C_B\leadsto C_A$, every path requires at least $K_1+1$
mistakes to first reach a state with $q=K_1+1$. It also requires at least
$K_2+1-n_1$ mistakes in subpopulation 2 to first reach $q=K_2+1$:
subpopulation 1 contributes at most $n_1$ agents choosing $A$, while
agents in subpopulation 2 prefer $B$ whenever $q\le K_2$. Therefore,
\[
r(C_B,C_A)\ge\max\{K_1+1,K_2+1-n_1\}.
\]

To attain this bound, set $h=\min\{K_1+1,n_2\}$.
Fig.~\ref{fig:state-path-to-A}
illustrates a path beginning
with mistaken switches to $A$ by $h$ agents in
subpopulation 2 and $K_1+1-h$ in subpopulation 1, reaching
$(K_1+1-h,h)$. Best-response switches in subpopulation 1 then lead
to $(n_1,h)$ while subpopulation 2 does not revise. After
$\max\{0,K_2+1-n_1-h\}$ further mistaken switches to $A$ in
subpopulation 2, $q\ge K_2+1$ and best responses lead to $C_A$.
Since $K_1+1\le n_1$ and $K_2+1-n_1\le n_2$, the path is feasible
and its total number of mistakes is
\[
\begin{aligned}
&K_1+1+\max\{0,K_2+1-n_1-h\}\\
&\qquad=\max\{K_1+1,K_2+1-n_1\}.
\end{aligned}
\]
This attains the lower bound and proves Eq.~\eqref{eq:r-BA-direct}.

For $C_A\leadsto C_B$, every path requires at least $N-K_2$
mistakes to first reach a state with $q=K_2$. It also requires at least
$n_1-K_1$ mistakes in subpopulation 1 to first reach $q=K_1$:
subpopulation 2 contributes at most $n_2$ of the $N-K_1$ agents choosing
$B$, while agents in subpopulation 1 prefer $A$ whenever $q>K_1$.
Therefore,
\[
r(C_A,C_B)\ge\max\{N-K_2,n_1-K_1\}.
\]

To attain this bound, set $g=\min\{N-K_2,n_1\}$.
Fig.~\ref{fig:state-path-to-B}
illustrates a path beginning
with mistaken switches to $B$ by $g$ agents in subpopulation 1 and
$N-K_2-g$ in subpopulation 2, reducing $q$ to $K_2$.
Best responses in subpopulation 2 then lead to $(n_1-g,0)$.
After $\max\{0,n_1-K_1-g\}$ further mistaken switches to $B$ in subpopulation 1,
$q\le K_1$ and best responses lead to $C_B$. These switches are
feasible since $N-K_2\le n_2$ and $K_1\ge0$, and the total number of mistakes is
\[
N-K_2+\max\{0,n_1-K_1-g\}
=\max\{N-K_2,n_1-K_1\}.
\]
This attains the lower bound and proves Eq.~\eqref{eq:r-AB-direct}.
\end{proof}

\begin{figure}[t]
\centering
\begin{tikzpicture}[
    >=Latex,
    node distance=2.15cm,
    state/.style={
        circle,
        draw,
        minimum size=10mm,
        font=\small
    },
    edge/.style={
        ->,
        thick,
        shorten >=2pt,
        shorten <=2pt
    },
    lab/.style={
        fill=white,
        inner sep=1pt,
        font=\scriptsize
    }
]
\node[state] (CB) {$C_B$};
\node[state, right=of CB] (P) {$P$};
\node[state, right=of P] (CA) {$C_A$};

% Left pair.
\draw[edge, bend left=20]
    (CB) to node[lab, above] {$K_1+1$} (P);
\draw[edge, bend left=20]
    (P) to node[lab, below] {$n_1-K_1$} (CB);

% Right pair.
\draw[edge, bend left=20]
    (P) to node[lab, above] {$K_2+1-n_1$} (CA);
\draw[edge, bend left=20]
    (CA) to node[lab, below] {$N-K_2$} (P);

% Transitions between C_B and C_A.
\draw[edge]
    (CB.north) .. controls +(0,1.5cm) and +(0,1.5cm) ..
    node[lab, above] {$\max\{K_1+1,\,K_2+1-n_1\}$} (CA.north);
\draw[edge]
    (CA.south) .. controls +(0,-1.5cm) and +(0,-1.5cm) ..
    node[lab, below] {$\max\{N-K_2,\,n_1-K_1\}$} (CB.south);
\end{tikzpicture}
\caption{
Minimum resistances between all three equilibria under the hypotheses of
Proposition~\ref{prop:two-type-minimum-resistances}.
Each arrow is labeled by the minimum number of mistakes required for
the indicated transition.
}
\label{fig:general-resistances}
\end{figure}
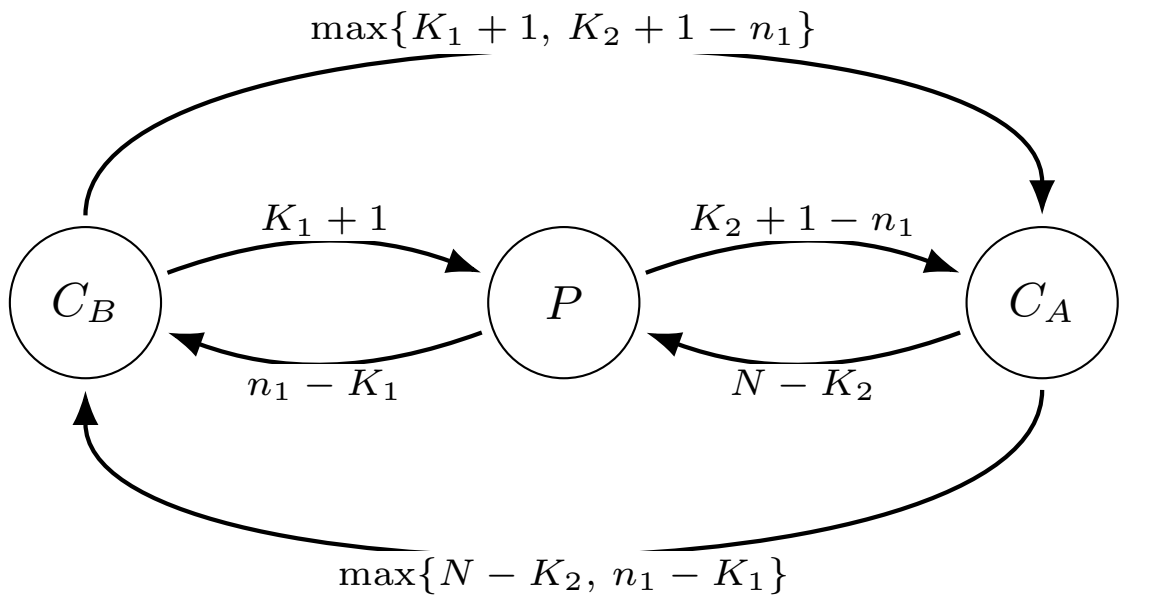

Fig.~\ref{fig:general-resistances} summarizes the minimum resistances
between all three equilibria in Proposition~\ref{prop:two-type-minimum-resistances}.

\section{Stochastic Stability and Welfare of Polarized Equilibria}
\label{sec:stochastic-selection}

In the previous section, we showed that heterogeneous risk attitudes alone can
give rise to a polarized equilibrium, even when agents share the same payoff
matrix and agree on the preferred coordinated outcome. We now ask whether
this coordination failure persists under risk-sensitive BRM dynamics and
whether the polarized equilibrium can become the uniquely selected long-run
outcome. Using the transition
resistances from Section~\ref{sec:learning-dynamics}, we characterize the
stochastic stability of the polarized equilibrium, interpret the conditions
for large populations, and examine the resulting welfare losses.

\subsection{Stochastic potentials and stability conditions}

Let $\mathcal C$ denote the set of risk-sensitive equilibria identified
in Proposition~\ref{prop:two-type-recurrent}, equivalently the absorbing
states of the Markov chain with transition matrix $M^0$. This set contains $C_B$ and
$C_A$, and also $P$ when $\tau_1<p<\tau_2$. Define the
\emph{reduced resistance graph} $\mathcal G_{\mathcal C}$ to have
vertex set $\mathcal C$ and an edge $u\to v$ of weight $r(u,v)$
for every pair of distinct equilibria. Thus, $\mathcal G^0$ records
individual best-response revisions on the full state space, whereas
$\mathcal G_{\mathcal C}$ records minimum-resistance paths between
equilibria. Following~\cite{young1993evolution}, we compare equilibria
using the total resistance of trees in $\mathcal G_{\mathcal C}$ directed
toward each equilibrium.
For $z\in\mathcal C$, a \emph{spanning tree directed toward $z$}
is a subgraph of $\mathcal G_{\mathcal C}$ containing all its vertices.
Each equilibrium other than $z$ has exactly one outgoing edge and a
directed path to $z$, while $z$ has no outgoing edge.
Its edges have weights $r(u,v)$. Let
$\mathcal T_z$ denote the set of these trees. We
define the \emph{stochastic potential} $\gamma_z$ of $z$ by
\[
\gamma_z:=\min_{T\in\mathcal T_z}\sum_{(u,v)\in T}r(u,v).
\]

\begin{proposition}[Resistance-tree selection, special case of {\cite[Theorem~4]{young1993evolution}}]
\label{prop:resistance-tree-selection}
Let $M^\varepsilon$ be a regular perturbation of $M^0$ on the finite
state space $\mathcal S$, and suppose the recurrent classes of $M^0$
are the singleton equilibria in $\mathcal C$. Then the stationary
distribution $\pi^\varepsilon$ converges as $\varepsilon\to0$, and its
limiting support consists exactly of the equilibria minimizing
$\gamma_z$. A unique minimizer has limiting probability one.
\end{proposition}

Propositions~\ref{prop:two-type-recurrent}
and~\ref{prop:mistake-resistance} verify the hypotheses of
Proposition~\ref{prop:resistance-tree-selection}. In particular, all transient states
have limiting stationary probability zero, so we identify the SSEs by
comparing the stochastic potentials of the risk-sensitive equilibria.
The next proposition gives these potentials when $P$ is an equilibrium.

\begin{proposition}[Stochastic potentials of the equilibria]
\label{prop:three-potentials}
Under the hypotheses of Proposition~\ref{prop:two-type-minimum-resistances},
write $\gamma_B:=\gamma_{C_B}$ and $\gamma_A:=\gamma_{C_A}$. Then, we have  
\small
\begin{align}
\gamma_P
&=
r(C_B,P)+r(C_A,P),
\label{eq:gamma-P}
\\
\gamma_B
&=
\min\Big\{
r(P,C_B)+r(C_A,P),
\nonumber\\[-1mm]
&\hspace{24mm}
r(P,C_A)+r(C_A,C_B)
\Big\},
\label{eq:gamma-B}
\\
\gamma_A
&=
\min\Big\{
r(C_B,P)+r(P,C_A),
\nonumber\\[-1mm]
&\hspace{24mm}
r(P,C_B)+r(C_B,C_A)
\Big\}.
\label{eq:gamma-A}
\end{align}
\end{proposition}
\normalsize

\begin{proof}
For root $P$, the tree with edges $C_B\to P$ and $C_A\to P$
has cost $r(C_B,P)+r(C_A,P)$. The tree $C_B\to C_A\to P$
has cost $r(C_B,C_A)+r(C_A,P)$, and the tree $C_A\to C_B\to P$
has cost $r(C_A,C_B)+r(C_B,P)$.
By Proposition~\ref{prop:two-type-minimum-resistances},
$r(C_B,C_A)\ge r(C_B,P)$ and
$r(C_A,C_B)\ge r(C_A,P)$. Hence, the first cost is minimal,
giving Eq.~\eqref{eq:gamma-P}.

For root $C_B$, the tree $C_A\to P\to C_B$ has cost
$r(C_A,P)+r(P,C_B)$. The tree $P\to C_A\to C_B$ has cost
$r(P,C_A)+r(C_A,C_B)$. The tree with edges $P\to C_B$ and
$C_A\to C_B$ has cost $r(P,C_B)+r(C_A,C_B)$, which is at least
the cost of $C_A\to P\to C_B$ because $r(C_A,C_B)\ge r(C_A,P)$.
Thus, the minimum of the first two costs gives Eq.~\eqref{eq:gamma-B}.

For root $C_A$, the tree $C_B\to P\to C_A$ has cost
$r(C_B,P)+r(P,C_A)$. The tree $P\to C_B\to C_A$ has cost
$r(P,C_B)+r(C_B,C_A)$. The tree with edges $C_B\to C_A$ and
$P\to C_A$ has cost $r(C_B,C_A)+r(P,C_A)$, which is at least
the cost of $C_B\to P\to C_A$ because $r(C_B,C_A)\ge r(C_B,P)$.
Thus, the minimum of the first two costs gives Eq.~\eqref{eq:gamma-A}.
\end{proof}

These expressions yield a particularly simple characterization of the
stochastic stability of the polarized equilibrium.

\begin{theorem}[Stochastically stable polarized equilibrium]
\label{thm:polarized-SSE}
Suppose Eq.~\eqref{eq:ordered-two-thresholds}, Assumption~\ref{ass:no-ties},
and $\tau_1<p<\tau_2$ hold.
Then, the polarized equilibrium
$P=(n_1,0)$ is stochastically stable if and only if
$r(C_B,P)\le r(P,C_B)$ and $r(C_A,P)\le r(P,C_A)$.
Equivalently,
\begin{equation}
2K_1+1
\le n_1
\le
2K_2+1-N.
\label{eq:finite-P-SSE}
\end{equation}
Moreover, $P$ is the unique SSE if and only if both inequalities
in Eq.~\eqref{eq:finite-P-SSE} are strict; in this case,
$\pi^\varepsilon(P)\to1$ as $\varepsilon\to0$.
\end{theorem}

\begin{proof}
Using Proposition~\ref{prop:two-type-minimum-resistances}, write
$u=r(C_B,P)=K_1+1$, $v=r(P,C_B)=n_1-K_1$,
$w=r(P,C_A)=K_2+1-n_1$, and $z=r(C_A,P)=N-K_2$.
Equations~\eqref{eq:r-AB-direct} and~\eqref{eq:r-BA-direct} then give
\[
\begin{aligned}
r(C_A,C_B)&=\max\{N-K_2,n_1-K_1\}=\max\{z,v\},\\
r(C_B,C_A)&=\max\{K_1+1,K_2+1-n_1\}=\max\{u,w\}.
\end{aligned}
\]
Substituting these expressions into Eqs.~\eqref{eq:gamma-P}--\eqref{eq:gamma-A} gives
\[
\begin{aligned}
\gamma_P&=u+z,\\
\gamma_B&=\min\{v+z,\ w+\max\{z,v\}\},\\
\gamma_A&=\min\{u+w,\ v+\max\{u,w\}\}.
\end{aligned}
\]
By Proposition~\ref{prop:resistance-tree-selection}, $P$ is an SSE
exactly when $\gamma_P\le\gamma_B$ and $\gamma_P\le\gamma_A$.

For necessity, if $u>v$, then
$\gamma_B\le v+z<u+z=\gamma_P$, so $P$ is not stochastically
stable. Likewise, if $z>w$, then
$\gamma_A\le u+w<u+z=\gamma_P$. Thus, stochastic stability
requires $u\le v$ and $z\le w$.
For sufficiency, suppose $u\le v$ and $z\le w$. The terms
$v+z$ and $u+w$ are both at least $u+z$.
The remaining terms in the minima defining $\gamma_B$ and $\gamma_A$ satisfy
\[
\begin{aligned}
w+\max\{z,v\}&\ge w+v\ge z+u,\\
v+\max\{u,w\}&\ge v+w\ge u+z.
\end{aligned}
\]
Hence, the potentials of $C_B$ and $C_A$ are at least $\gamma_P$, proving
the claimed equivalence. Substituting the minimum resistances yields
$K_1+1\le n_1-K_1$ and $N-K_2\le K_2+1-n_1$,
which can be rearranged  to obtain Eq.~\eqref{eq:finite-P-SSE}.

If $u<v$ and $z<w$, the four terms $v+z$, $w+\max\{z,v\}$,
$u+w$, and $v+\max\{u,w\}$ in the minima defining $\gamma_B$ and
$\gamma_A$ all strictly exceed $u+z=\gamma_P$. Thus, $P$ uniquely minimizes the
stochastic potential, so $\pi^\varepsilon(P)\to1$ by
Proposition~\ref{prop:resistance-tree-selection}.
If instead $u=v$ while $z\le w$, then $\gamma_B=\gamma_P$;
if $z=w$ while $u\le v$, then $\gamma_A=\gamma_P$.
In the first case, both $P$ and $C_B$ are stochastically stable; in
the second, both $P$ and $C_A$ are stochastically stable. Thus, $P$
is an SSE but is not uniquely selected in either equality case.
\end{proof}

The polarized equilibrium $P$ is stochastically stable only if each group
accounts for a sufficiently large share of the population. The left inequality
in Eq.~\eqref{eq:finite-P-SSE} requires $n_1\ge 2K_1+1$ for subpopulation~1.
Since $n_2=N-n_1$, the right inequality requires $n_2\ge 2N-2K_2-1$ for
subpopulation~2. If either group falls below its bound, the stationary
probability of $P$ vanishes as the mutation probability $\varepsilon$
tends to zero. If both inequalities are strict, $P$ is the unique
stochastically stable equilibrium, even though every agent would benefit
from coordination on $A$.

\subsection{Large-population interpretation}

We now let the population size $N$ grow while the population share $p$ and
the risk parameters $\beta_1,\beta_2$ remain fixed.
The selection criterion then takes the following limiting form.

\begin{corollary}[]
\label{cor:large-P-SSE}
Suppose Eq.~\eqref{eq:ordered-two-thresholds} and $\tau_1<p<\tau_2$ hold,
and let $N\to\infty$ along population sizes that satisfy
Assumption~\ref{ass:no-ties} and $pN\in\mathbb Z$.
Then the criterion in Theorem~\ref{thm:polarized-SSE} reduces to
\begin{equation}
2\tau_1
\le p
\le
2\tau_2-1.
\label{eq:large-P-SSE}
\end{equation}
\end{corollary}

\begin{proof}
Since $K_k=\lfloor N\tau_k\rfloor$ satisfies $N\tau_k-1<K_k\le N\tau_k$,
we have $K_k/N\to\tau_k$ for $k\in\{1,2\}$. Dividing the criterion
$2K_1+1\le n_1\le2K_2+1-N$ in Eq.~\eqref{eq:finite-P-SSE} by $N$, using
$n_1=pN$, and letting $N\to\infty$ gives Eq.~\eqref{eq:large-P-SSE}.
\end{proof}

The strict bounds in Eq.~\eqref{eq:large-P-SSE} agree with those obtained
from~\cite[Theorem~3]{neary2012competing} after normalizing the integer
best-response thresholds by population size and taking the large-population
limit. That theorem assumes deterministic group adjustment at equal,
constant rates: each period, each group switches a fixed number of agents
to its best response, capped by the number still using the other action.
Both groups best respond to the same pre-update population state, after which
agents mutate independently. Our asynchronous BRM process selects one
agent uniformly at random to best respond or mutate at each step.
The source of heterogeneity also differs. Agents in our model share a
common payoff matrix and differ only in their risk attitudes, whereas
Neary's subpopulations have different payoff matrices. The welfare
implications of this distinction are discussed after
Proposition~\ref{prop:pareto-inefficient-sse}.

The condition in Eq.~\eqref{eq:large-P-SSE} is stronger than the
equilibrium condition $\tau_1<p<\tau_2$ in
Proposition~\ref{prop:two-type-recurrent}: it requires $p$ to lie
sufficiently far from both best-response thresholds. Specifically, $p$
must exceed $\tau_1$ by at least $\tau_1$ and fall below $\tau_2$ by at
least $1-\tau_2$. Such a $p$ can exist only if the two thresholds are
themselves sufficiently far apart, which has an immediate implication for
the heterogeneity in risk attitudes between the two subpopulations.

\begin{corollary}[Threshold separation]
\label{cor:threshold-separation}
There exists a population composition $p$ for which the polarized equilibrium is
the unique SSE when the population is sufficiently large if and only if
$\tau_2-\tau_1>1/2$.
When this condition holds, any $p\in(2\tau_1,\,2\tau_2-1)$ makes the
polarized equilibrium the unique SSE for all sufficiently large populations.
\end{corollary}

Thus, when the two subpopulations have sufficiently separated
best-response thresholds and each subpopulation's share is large enough
that $p$ lies in the interval $(2\tau_1,\,2\tau_2-1)$, the polarized
equilibrium is uniquely stochastically stable in sufficiently large
populations.
Since the threshold map $\tau(\cdot)$ is strictly decreasing, its inverse
$h:=\tau^{-1}$ is also strictly decreasing, and the condition
$2\tau_1<p<2\tau_2-1$ is equivalent to
$\beta_1>h(p/2)$ and $\beta_2<h((1+p)/2)$.
Hence, when the polarized equilibrium is the unique SSE, agents in
subpopulation 1 are sufficiently risk-seeking ($\beta_1>0$), while the
risk parameter $\beta_2$ lies below the cutoff $h((1+p)/2)$ but need not
be negative.

\subsection{Example: Inefficient polarization under risk heterogeneity}

To illustrate the large-population condition in
Eq.~\eqref{eq:large-P-SSE}, we return to the game
in Eq.~\eqref{eq:motivating-game} introduced in
Section~\ref{subsec:motivating-example}, which satisfies
Assumption~\ref{ass:payoffs}. We again take $\beta_1=2$, $\beta_2=-0.5$,
and $p=0.30$; recall that the best-response thresholds are then
$\tau_1\approx0.0204$ and $\tau_2\approx0.8780$. Since
$2\tau_1\approx0.0407<p$ and $p<0.7561\approx2\tau_2-1$,
Corollary~\ref{cor:threshold-separation} implies that $P$ is the unique
SSE for all sufficiently large populations.

For $N=100$, we have $n_1=30$, $K_1=2$, and $K_2=87$.
The minimum resistances between $P$ and each of $C_B$ and $C_A$ are
$r(C_B,P)=3$, $r(P,C_B)=28$, $r(C_A,P)=13$, and $r(P,C_A)=58$.
Thus, the polarized equilibrium is substantially easier to enter than to leave
from either direction. The stochastic potentials are $\gamma_B=41$,
$\gamma_P=16$, and $\gamma_A=61$, which implies that $P$ is the unique
SSE and $\pi^\varepsilon(P)\to1$ as $\varepsilon\to 0$.

Every agent nevertheless agrees that certain coordination on $A$
yields the larger payoff. The welfare comparison is
\[
W(P)=1.97<2=W(C_B)<4=W(C_A).
\]
Hence, risk heterogeneity uniquely selects a polarized
equilibrium with lower welfare than either $C_B$ or $C_A$.
The next subsection characterizes these welfare comparisons and shows that coordination on $A$ improves every agent's payoff.

\subsection{Pareto inefficiency of the stochastically stable polarized equilibrium}
\label{subsec:pareto-polarized-equilibria}

We compare the population states $P$ and $C_A$ using two welfare
criteria: each agent's \emph{expected payoff} from random matching and
her own ERM-adjusted payoff. The equilibrium condition for $P$ uses
ERM-adjusted best responses, as in Definition~\ref{def:risk-sensitive-equilibrium}.
The payoff-dominance label for $C_A$ refers to the underlying two-player
game. The following result shows that $C_A$ strictly Pareto dominates $P$,
even when $P$ is a stochastically stable risk-sensitive equilibrium.
We then characterize when welfare $W(P)$ also falls below
$W(C_B)$.

\begin{proposition}[Pareto-inefficient polarized equilibria]
\label{prop:pareto-inefficient-sse}
Under Assumption~\ref{ass:payoffs} (\ref{ass:payoff-dominance}), 
$C_A$ strictly Pareto dominates $P$
under both expected payoffs and each agent's own ERM-adjusted
payoffs. Hence, every polarized SSE characterized by
Theorem~\ref{thm:polarized-SSE} is Pareto inefficient in both senses.
\end{proposition}

\begin{proof}
Since $a>c$, $d>b$, and $a>d$ by
Assumption~\ref{ass:payoffs}(\ref{ass:payoff-dominance}), $a$ is the strictly
largest payoff, and every agent receives it with certainty at $C_A$.
At $P$, agents in subpopulations 1 and 2 receive $Z_A(p)$ and $Z_B(p)$,
respectively; since $0<p<1$, each such payoff $Z$ satisfies $Z\le a$ and
$\Pr(Z<a)>0$. It therefore suffices to show that $\ERM_\beta(Z)<a$ for every
$\beta\in\mathbb R$, where $\beta=0$ gives expected payoffs.
For $\beta=0$, this is $\mathbb E[Z]<a$. For $\beta\neq0$, we have
$\mathbb E[e^{\beta Z}]<e^{\beta a}$ if $\beta>0$ and
$\mathbb E[e^{\beta Z}]>e^{\beta a}$ if $\beta<0$; taking logarithms and
dividing by $\beta$, which reverses the inequality when $\beta<0$, gives
$\ERM_\beta(Z)<a$ in both cases.
\end{proof}

In~\cite[Theorem~7]{neary2012competing}, neither equilibrium in which all agents choose the same action
Pareto dominates a polarized SSE. Here, $C_A$ strictly Pareto dominates
the polarized equilibrium under both expected payoffs and each agent's
ERM-adjusted payoffs. This shows that risk heterogeneity alone under an identical payoff matrix
can have an inefficient polarized equilibrium emerge despite unanimous agreement on the
preferred state $C_A$.

\begin{proposition}[Welfare of stochastically stable polarization]
\label{prop:welfare-polarized-sse}
Under Assumption~\ref{ass:payoffs}, suppose the hypotheses of
Theorem~\ref{thm:polarized-SSE} hold. Let $\Delta:=a-b-c+d>0$ and define
\begin{equation}
p_B:=\frac{2d-b-c}{a-b-c+d}
=1-\frac{a-d}{\Delta}.
\label{eq:polarized-welfare-cutoff}
\end{equation}
Then $W(P)<W(C_A)=a$ for every $p\in(0,1)$, while
\begin{equation}
W(P)<W(C_B)=d
\quad\Longleftrightarrow\quad p<p_B.
\label{eq:polarized-welfare-below-B}
\end{equation}
In particular, $p_B\in(0,1)$ if and only if $b+c<2d$.

For fixed risk parameters and population share $p$, a sufficient condition
for $P$ to be the unique SSE with welfare strictly below both coordinated
states in all sufficiently large populations satisfying
Assumption~\ref{ass:no-ties} and $pN\in\mathbb Z$ is
\begin{equation}
2\tau_1<p<\min\{2\tau_2-1,\,p_B\}.
\label{eq:large-welfare-inferior-P-SSE}
\end{equation}
\end{proposition}

\begin{proof}
Since $\Delta=(a-c)+(d-b)>0$, Eq.~\eqref{eq:welfare} gives
\begin{align*}
W(C_A)-W(P)&=(1-p)(a-d+\Delta p)>0,\\
W(P)-W(C_B)&=\Delta p(p-p_B).
\end{align*}
These identities establish the welfare comparisons. Since $a>d$,
$p_B<1$, while $p_B>0$ if and only if $b+c<2d$.
The large-population conclusion follows by combining $p<p_B$ with
Corollary~\ref{cor:threshold-separation}.
\end{proof}

When $b+c<2d$, the cutoff $p_B$ is the share choosing $A$ at which
aggregate welfare recovers the level under coordination on $B$; smaller
positive shares yield lower welfare. This cutoff depends only on the
payoff matrix, while risk attitudes determine whether polarization at a
given share is stochastically stable. In particular, if
$2\tau_1<2\tau_2-1\le p_B$, every share in the open large-population
region where $P$ is the unique SSE yields lower welfare than $C_B$.

This inefficiency motivates interventions that steer long-run behavior away
from both the inefficient coordinated outcome $C_B$ and inefficient
polarization toward $C_A$, where every agent is better off. We therefore
ask: \emph{what fraction of agents must have their risk attitudes
changed, and by how much, to make $C_A$ the unique stochastically stable
equilibrium?}

\section{Interventions on Risk Attitudes for Efficient Stochastic Selection}
\label{sec:intervention}

Based on the results above, a planner can influence stochastic equilibrium
selection under agents' learning dynamics by choosing a subset of agents
and modifying their risk attitudes. This is particularly relevant for
\emph{autonomous agents} that can be controlled or trained by human designers.
The intervention modifies only agents' risk attitudes and requires neither
changes to the payoff matrix nor direct constraints on agents' actions.
We study how large a fraction of agents must have their risk attitudes
modified (the \emph{intervention share}) and by how much each modified agent's
risk parameter must increase (the \emph{intervention intensity}) to make the
payoff-dominant equilibrium $C_A$ the unique SSE.

Initially, suppose all $N$ agents have risk parameter $\beta_L$ and threshold
$\tau_L=\tau(\beta_L)>1/2$. In this case, $C_B$ is an SSE before intervention.
We focus on this regime because it is the only one requiring an intervention
share bounded away from zero in the large-population limit. If $\tau_L<1/2$,
$C_A$ is already the unique SSE for all sufficiently large $N$, so no intervention
is needed.\footnote{For a finite homogeneous population, $C_A$ is the
unique SSE exactly when $2\lfloor N\tau_L\rfloor+1<N$; equality makes both $C_A$
and $C_B$ SSEs. The finite-population criterion in Corollary~\ref{cor:finite-control}
accounts for these ties near $\tau_L=1/2$. At the boundary $\tau_L=1/2$,
lowering one agent's threshold to any fixed $\tau_H<1/2$ suffices to make $C_A$
the unique SSE for all sufficiently large admissible populations, so the
required intervention share is $1/N\to0$.}
The planner then chooses any subset of $n<N$ agents from this
initially homogeneous population and raises their risk parameter to
$\beta_H>\beta_L$, leaving the remaining $N-n$ agents unchanged.
The intervention thus creates the two subpopulations, with the modified
agents forming subpopulation 1 and the remaining agents forming
subpopulation 2. The planner controls the \emph{intervention share}
$p=n/N$ (the share of subpopulation 1, as before, since $n_1=n$)
and \emph{intervention intensity} $\Delta\beta:=\beta_H-\beta_L>0$.
Since $\tau(\cdot)$ is strictly decreasing in the risk parameter $\beta$ under Assumption~\ref{ass:payoffs},
greater intervention intensity at a fixed $\beta_L$ means greater risk seeking and a lower
threshold $\tau_H=\tau(\beta_H)<\tau_L$.
We next derive conditions on the intervention share $p$ and intervention intensity
$\Delta\beta$ under which the payoff-dominant equilibrium $C_A$ is the unique SSE.
For a fixed initial risk parameter $\beta_L$, we express these conditions using
the corresponding thresholds $\tau_L=\tau(\beta_L)$ and
$\tau_H=\tau(\beta_L+\Delta\beta)$.

\subsection{Conditions for efficient stochastic selection}\label{sec:cond_efficient}

Following Section~\ref{subsec:population-states}, write the integer thresholds as
$K_H:=\lfloor N\tau_H\rfloor$ and $K_L:=\lfloor N\tau_L\rfloor$, with subscripts
matching the risk parameters, and maintain
$N\tau_H,N\tau_L\notin\mathbb Z$ as required by Assumption~\ref{ass:no-ties},
with $\tau_1=\tau_H$ and $\tau_2=\tau_L$. Applying the
resistance-tree criterion in Proposition~\ref{prop:resistance-tree-selection}
to these two subpopulations gives the following corollary.

\begin{corollary}[Finite-population condition for efficient selection]
\label{cor:finite-control}
Suppose $0<\tau_H<\tau_L<1$, with $n$ agents having threshold $\tau_H$ and the remaining
$N-n$ having threshold $\tau_L$. Then, $C_A$ is the unique SSE if and only if
\[
\begin{aligned}
n&>2K_L+1-N,\\
K_H+1&<\max\{N-K_L,\,n-K_H\}.
\end{aligned}
\]
\end{corollary}

\begin{proof}
Set $u=K_H+1$, $v=n-K_H$, $w=K_L+1-n$, and $z=N-K_L$.
By Proposition~\ref{prop:resistance-tree-selection}, $C_A$ is the unique SSE
exactly when its stochastic potential is strictly smaller than every competing potential.

If $K_H<n\le K_L$, the equilibria are $C_B$, $P$, and $C_A$.
Propositions~\ref{prop:two-type-minimum-resistances} and~\ref{prop:three-potentials}
give
\[
\begin{aligned}
\gamma_P&=u+z,\\
\gamma_B&=\min\{v+z,\,w+\max\{z,v\}\},\\
\gamma_A&=\min\{u+w,\,v+\max\{u,w\}\}.
\end{aligned}
\]
If $w\ge z$, these formulas imply $\gamma_A\ge\min\{\gamma_P,\gamma_B\}$,
excluding unique selection of $C_A$.
If $w<z$, then $\gamma_A<\gamma_P$, while $\gamma_A<\gamma_B$ is equivalent
to $u<v$ when $v>z$ and to $u<z$ when $v\le z$.
Thus, $C_A$ is the unique SSE exactly when $w<z$ and $u<\max\{z,v\}$.

Otherwise, $P$ is absent and only $C_B$ and $C_A$ remain by
Proposition~\ref{prop:two-type-recurrent}. Each resistance tree has one edge,
so the threshold-crossing argument for Eqs.~\eqref{eq:r-BA-direct}
and~\eqref{eq:r-AB-direct} gives
\[
\begin{aligned}
\gamma_A&=r(C_B,C_A)=\max\{u,w\},\\
\gamma_B&=r(C_A,C_B)=\max\{z,v\}.
\end{aligned}
\]
For $n\le K_H$, we have $v\le0<z$, so $\gamma_A<\gamma_B$ requires
exactly $u<z$ and $w<z$.
For $n\ge K_L+1$, we have $w\le0<u$, so it requires exactly
$u<\max\{z,v\}$, with $w<z$ automatic.
Hence, the same criterion holds whether or not $P$ exists.

Substituting the definitions into $w<z$ and $u<\max\{z,v\}$ gives the stated conditions.
\end{proof}

The condition $n>2K_L+1-N$ requires a sufficiently large intervention share,
while $K_H+1<\max\{N-K_L,\,n-K_H\}$ requires sufficiently high intervention
intensity for the chosen intervention share. The following corollary expresses
these two conditions in terms of $p$ and the corresponding threshold $\tau_H$
in the large-population limit.

\begin{corollary}[Efficient stochastic-selection region]
\label{cor:control-region}
Let $N\to\infty$ with $p$, $\tau_H$, and $\tau_L$ fixed, along population
sizes satisfying Assumption~\ref{ass:no-ties} and $pN\in\mathbb Z$.
Then, $C_A$ is the unique SSE for all sufficiently large $N$ if both
of the following conditions hold:
\begin{align}
p&>2\tau_L-1
\quad\text{(intervention share condition)},
\label{eq:intervention-share-condition}\\
\tau_H&<\max\{1-\tau_L,\,p/2\}
\notag\\[-0.5ex]
&\hspace{2em}\text{(intervention intensity condition)}.
\label{eq:intervention-intensity-condition}
\end{align}
\end{corollary}

\begin{proof}
Divide the inequalities in Corollary~\ref{cor:finite-control} by $N$ and use
$K_H/N\to\tau_H$ and $K_L/N\to\tau_L$. The limiting inequalities are
$p>2\tau_L-1$ and $\tau_H<\max\{1-\tau_L,\,p-\tau_H\}$, the latter being
equivalent to Eq.~\eqref{eq:intervention-intensity-condition}.
Their strictness ensures that the finite-population inequalities hold for all
sufficiently large admissible $N$.
\end{proof}

The intervention share condition~\eqref{eq:intervention-share-condition} sets
a lower bound on the fraction of agents whose risk attitudes must be modified.
The intervention intensity condition~\eqref{eq:intervention-intensity-condition}
requires sufficiently high $\Delta\beta$ that the corresponding threshold
$\tau_H$ falls below $\tau_c(p):=\max\{1-\tau_L,p/2\}$.
Since $\tau_L=\tau(\beta_L)$ and $\tau_H=\tau(\beta_L+\Delta\beta)$,
Eqs.~\eqref{eq:intervention-share-condition} and~\eqref{eq:intervention-intensity-condition}
can be written as $p>2\tau(\beta_L)-1$ and
$\Delta\beta>h(\tau_c(p))-\beta_L$, respectively, where $h(\cdot)=\tau^{-1}(\cdot)$
is the inverse threshold map introduced in Section~\ref{sec:stochastic-selection}.

If $\tau_H<1-\tau_L$, the intervention intensity condition~\eqref{eq:intervention-intensity-condition}
holds for every intervention share. Thus, only the intervention share
condition~\eqref{eq:intervention-share-condition} remains to be met.
The intervention share must exceed $2\tau_L-1$, regardless of the intervention
intensity. In other words, a planner who controls too small a share of agents cannot make $C_A$ the unique
SSE simply by making those agents more risk-seeking.

Combining Eqs.~\eqref{eq:intervention-share-condition} and~\eqref{eq:intervention-intensity-condition}
gives the infimum intervention share required for efficient selection at a fixed
intervention intensity $\Delta\beta$, expressed through its corresponding threshold
$\tau_H=\tau(\beta_L+\Delta\beta)$:
\begin{equation}
p_c(\tau_H)=
\begin{cases}
2\tau_L-1, & \tau_H<1-\tau_L,\\
\max\{2\tau_L-1,\,2\tau_H\}, & 1-\tau_L\le \tau_H<1/2.
\end{cases}
\label{eq:share-frontier}
\end{equation}
For $\tau_H<1/2$, Eq.~\eqref{eq:share-frontier} summarizes the two conditions
as $p>p_c(\tau_H)$.
If $\tau_H\ge1/2$, the intervention intensity condition~\eqref{eq:intervention-intensity-condition}
cannot hold for any $p\le1$. Thus, the intervention intensity must exceed
$h(1/2)-\beta_L$ so that $\tau_H<1/2$. In other words, if the planner cannot make the modified agents sufficiently
risk-seeking, influencing more agents cannot make $C_A$ the unique SSE.

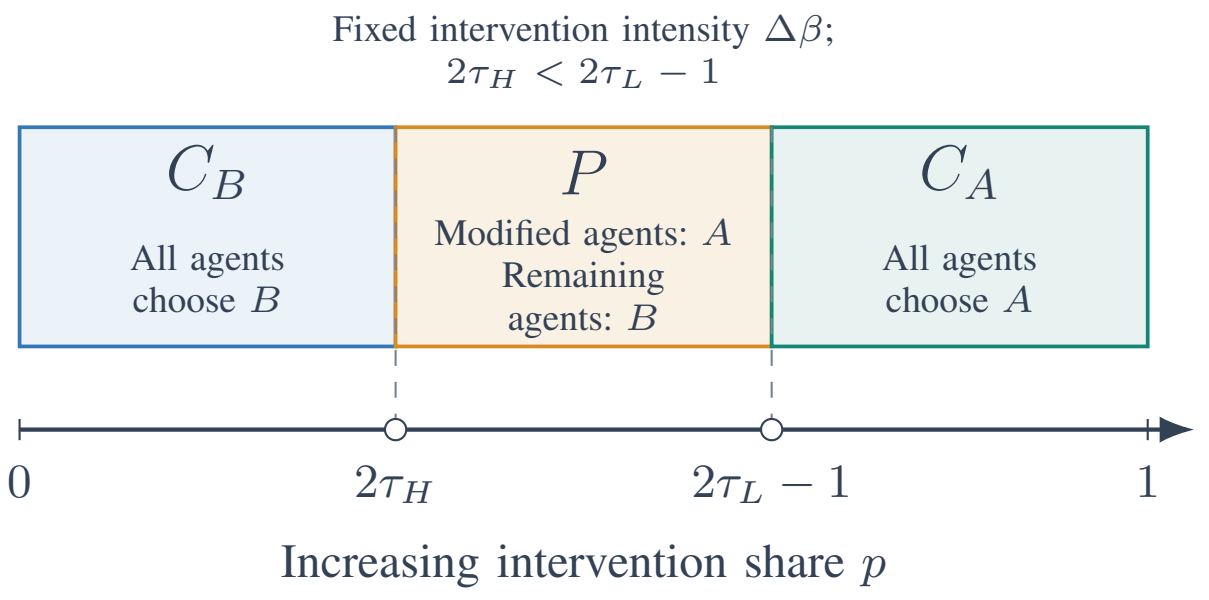
\begin{figure}[htbp]
\centering
\begingroup
\definecolor{shareCoordinatedB}{HTML}{3478B8}
\definecolor{sharePolarized}{HTML}{D88A25}
\definecolor{shareCoordinatedA}{HTML}{168576}
\definecolor{shareInk}{HTML}{334155}
\begin{tikzpicture}[
    x=0.072\linewidth, y=1cm,
    >=Latex,
    region/.style={line width=0.7pt},
    outcome/.style={font=\large, text=shareInk},
    actions/.style={font=\scriptsize, align=center, text width=0.25\linewidth, text=shareInk},
    tick label/.style={font=\small, text=shareInk, anchor=north}
]
\node[font=\scriptsize, align=center, text=shareInk] at (6,2.5)
    {Fixed intervention intensity $\Delta\beta$;\\$2\tau_H<2\tau_L-1$};

\path[region, draw=shareCoordinatedB, fill=shareCoordinatedB!10]
    (0,0.55) rectangle (4,2.0);
\path[region, draw=sharePolarized, fill=sharePolarized!12]
    (4,0.55) rectangle (8,2.0);
\path[region, draw=shareCoordinatedA, fill=shareCoordinatedA!10]
    (8,0.55) rectangle (12,2.0);
\node[outcome] at (2,1.7) {$C_B$};
\node[outcome] at (6,1.7) {$P$};
\node[outcome] at (10,1.7) {$C_A$};
\node[actions] at (2,1.0) {All agents\\choose $B$};
\node[actions] at (6,1.0) {Modified agents: $A$\\Remaining agents: $B$};
\node[actions] at (10,1.0) {All agents\\choose $A$};

\draw[->, line width=0.8pt, draw=shareInk] (0,0) -- (12.5,0);
\foreach \x in {0,12} {
    \draw[draw=shareInk] (\x,-0.07) -- (\x,0.07);
}
\foreach \x in {4,8} {
    \draw[dashed, draw=shareInk!65] (\x,0) -- (\x,2.0);
    \draw[draw=shareInk, fill=white] (\x,0) circle[radius=2pt];
}
\node[tick label] at (0,-0.12) {$0$};
\node[tick label] at (4,-0.12) {$2\tau_H$};
\node[tick label] at (8,-0.12) {$2\tau_L-1$};
\node[tick label] at (12,-0.12) {$1$};
\node[font=\small, text=shareInk] at (6,-0.9)
    {Increasing intervention share $p$};
\end{tikzpicture}
\endgroup
\caption{Selection as the planner increases the intervention share $p$ at fixed
intervention intensity $\Delta\beta$ and initial risk parameter $\beta_L$,
with $2\tau_H<2\tau_L-1$. Each interval shows the unique SSE for
fixed $p$ and sufficiently large admissible $N$.}
\label{fig:intervention-share-selection}
\end{figure}

Fix the initial risk parameter $\beta_L$ and intervention intensity $\Delta\beta$,
so the thresholds $\tau_L$ and $\tau_H$ remain fixed, and suppose
$2\tau_H<2\tau_L-1$.
As the planner increases the intervention share $p$, the selected outcome changes
across the three regions in Fig.~\ref{fig:intervention-share-selection}.
For all sufficiently large admissible populations,
$C_B$ is the unique SSE when $p<2\tau_H$: both groups coordinate on $B$.
When $2\tau_H<p<2\tau_L-1$, $P$ is the unique SSE: the modified agents choose
$A$, while the remaining agents choose $B$.
Once $p>2\tau_L-1$, $C_A$ is the unique SSE, and both groups coordinate on $A$.
Thus, increasing the intervention share can replace
$C_B$ with persistent polarization without achieving efficient coordination.
The intermediate outcome $P$ remains strictly Pareto dominated by $C_A$
(Proposition~\ref{prop:pareto-inefficient-sse}). It can also yield lower welfare
than $C_B$, as illustrated in Section~\ref{subsec:intervention-example}.
The planner must raise the intervention share $p$ beyond
the entire polarized region to achieve unique selection of $C_A$.

\subsection{Modifying risk attitudes versus adding new agents}
\label{subsec:modifying-versus-adding}

So far, we have considered modifying the risk attitudes of agents already in
the population. An alternative is to add new, more risk-seeking agents with
threshold $\tau_H<\tau_L$, while leaving the original agents' risk attitudes
unchanged. Both approaches increase the intervention share $p$ of agents with
threshold $\tau_H$. For added agents, the intervention intensity
$\Delta\beta=\beta_H-\beta_L$ measures their risk-parameter difference from
the original agents.
We now compare how many agents must be modified or added to make $C_A$ the
unique SSE.

We retain the initial setting from Section~\ref{sec:intervention}: a homogeneous
population of $N_0$ agents with threshold $\tau_L>1/2$. As discussed there, this
is the regime requiring an intervention share bounded away from zero in the
large-population limit. The resistance-tree criterion in
Proposition~\ref{prop:resistance-tree-selection} implies that $C_B$ is the unique SSE
before intervention for all sufficiently large admissible $N_0$.
For the comparison, fix the same intervention intensity $\Delta\beta$ for both
approaches, with $\tau_H=\tau(\beta_L+\Delta\beta)<1-\tau_L$ for the modified
or added agents. This satisfies the intervention intensity
condition~\eqref{eq:intervention-intensity-condition}, so efficient selection
in the large-population limit is governed by the intervention share
condition~\eqref{eq:intervention-share-condition}.

The planner first considers working with the existing population.
Modifying the risk attitudes of $m$ agents gives an intervention share
$p=m/N_0$ while keeping the population size fixed. Substituting this share into
Eq.~\eqref{eq:intervention-share-condition} gives the required number of modified
agents: $m>N_0(2\tau_L-1)$.
Now suppose the planner leaves the existing agents' risk attitudes
unmodified and adds $M$ agents with threshold $\tau_H$. These agents join a larger
population of size $N_0+M$ and form subpopulation 1, so the intervention share is
$p=M/(N_0+M)$.
Substituting this share into Eq.~\eqref{eq:intervention-share-condition} gives
the required number of added agents: $M>N_0(2\tau_L-1)/[2(1-\tau_L)]$.

The lower bound on $M$ is larger than the lower bound on $m$ by a factor of
$1/[2(1-\tau_L)]>1$, since $\tau_L>1/2$.
An intervention of intensity $\Delta\beta$ on an existing agent lowers her
threshold from $\tau_L$ to $\tau_H$. There is then one fewer agent with threshold $\tau_L$
and one more with threshold $\tau_H$, while the total population size stays fixed.
Every new agent added under the second approach has threshold $\tau_H$ and
therefore joins the group of more risk-seeking agents. However, all $N_0$
original agents retain threshold $\tau_L$, and the total population grows.
Adding the same number of agents therefore produces a smaller intervention share
than modifying that number of existing agents at the same intervention intensity.
This explains why modifying existing agents requires the planner to influence
fewer agents to achieve efficient selection in the large-population limit.

\subsection{Example: Modifying existing agents' risk attitudes versus adding risk-seeking agents}
\label{subsec:intervention-example}

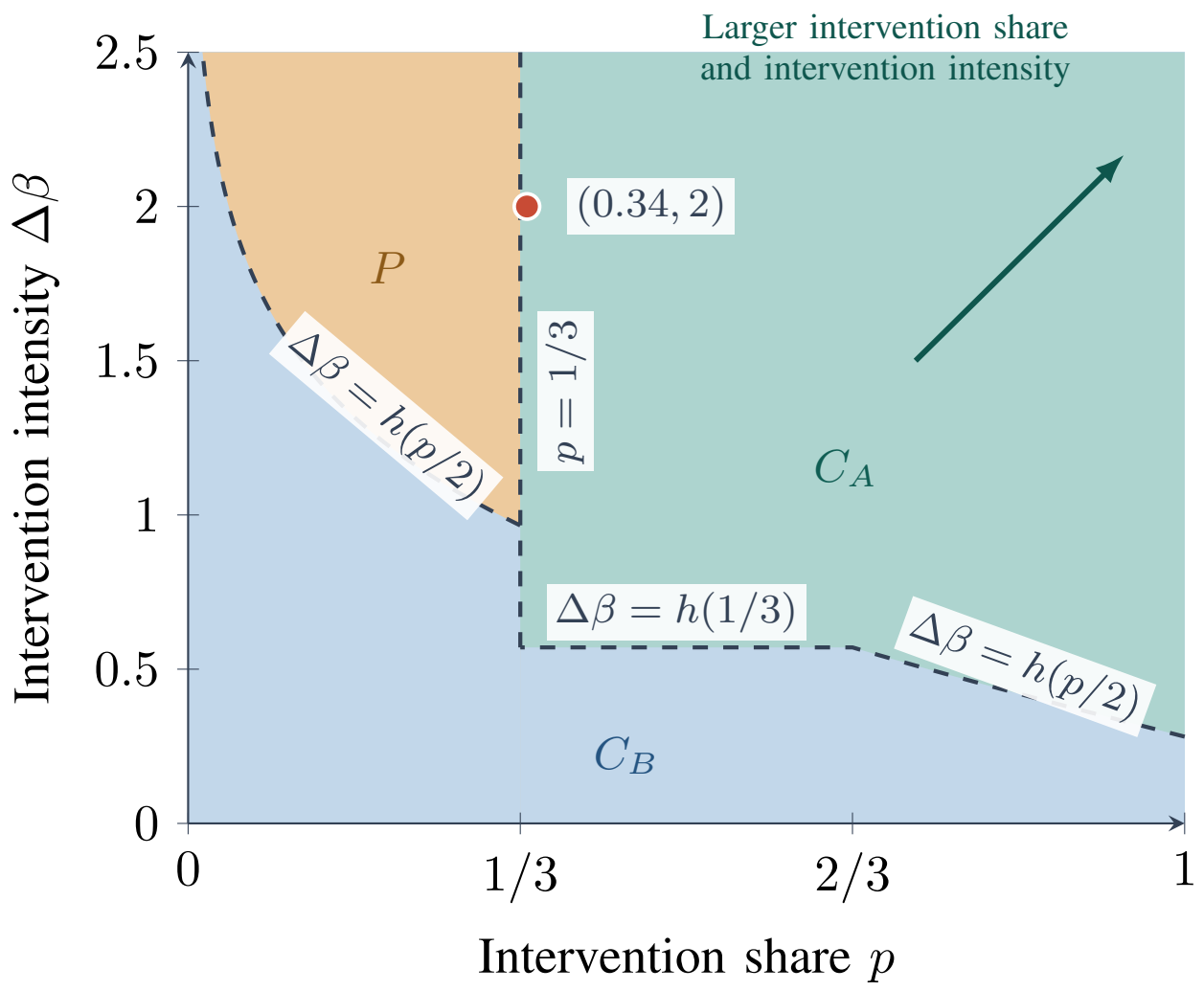
\begin{figure}[t]
\centering

% For this game and beta_L=0, beta_H equals intervention intensity.
% The constants below solve tau(Delta beta)=1/2, 1/3, and 1/6.
\def\intensityHalf{0.2811995743}
\def\intensityThird{0.5705796668}
\def\intensitySixth{0.9659814751}
\def\intensityMax{2.5}
\pgfmathsetmacro{\pc}{1/3}
\definecolor{coordinatedB}{HTML}{3478B8}
\definecolor{polarizedEquilibrium}{HTML}{D88A25}
\definecolor{coordinatedA}{HTML}{168576}
\definecolor{examplePoint}{HTML}{C84B35}
\definecolor{diagramInk}{HTML}{334155}

\begin{tikzpicture}[
    declare function={
        interventionTau(\b)=(exp(-2*\b)+exp(-3*\b))
            /(1+exp(-2*\b)+exp(-3*\b));
    },
    intervention label/.style={
        fill=white, fill opacity=0.9, text opacity=1,
        text=diagramInk, inner sep=1.5pt, font=\footnotesize
    },
    region label/.style={font=\small, inner sep=1pt}
]
\begin{axis}[
    width=0.98\columnwidth,
    height=0.80\columnwidth,
    xmin=0, xmax=1,
    ymin=0, ymax=\intensityMax,
    axis lines=left,
    axis on top,
    xlabel={Intervention share $p$},
    ylabel={Intervention intensity $\Delta\beta$},
    axis line style={diagramInk},
    tick style={diagramInk},
    xtick={0,0.3333333,0.6666667,1},
    xticklabels={$0$,$1/3$,$2/3$,$1$},
    ytick={0,0.5,1,1.5,2,2.5},
    clip=false,
]

% C_B occupies the background; the other regions lie above their boundaries.
\path[fill=coordinatedB!30]
    (axis cs:0,0) rectangle (axis cs:1,\intensityMax);

% P: 2*tau(Delta beta) < p < 1/3.
\addplot[
    draw=none, fill=polarizedEquilibrium!45,
    domain=\intensitySixth:\intensityMax, samples=120
] ({2*interventionTau(x)}, {x})
    -- (axis cs:\pc,\intensityMax) \closedcycle;

% C_A: p > 1/3 and Delta beta > h(max{1/3,p/2}).
\addplot[
    draw=none, fill=coordinatedA!35,
    domain=\intensityHalf:\intensityThird, samples=80
] ({2*interventionTau(x)}, {x})
    -- (axis cs:\pc,\intensityThird)
    -- (axis cs:\pc,\intensityMax)
    -- (axis cs:1,\intensityMax) \closedcycle;

% Selection boundaries in intervention share and intervention intensity.
\addplot[
    draw=diagramInk, dashed, thick,
    domain=\intensitySixth:\intensityMax, samples=120
] ({2*interventionTau(x)}, {x});
\addplot[
    draw=diagramInk, dashed, thick,
    domain=\intensityHalf:\intensityThird, samples=80
] ({2*interventionTau(x)}, {x})
    -- (axis cs:\pc,\intensityThird);
\addplot[draw=diagramInk, dashed, thick]
    coordinates {(\pc,\intensityThird) (\pc,\intensityMax)};

\node[region label, text=polarizedEquilibrium!65!black]
    at (axis cs:0.20,1.80) {$P$};
\node[region label, text=coordinatedA!70!black]
    at (axis cs:0.66,1.15) {$C_A$};
\node[region label, text=coordinatedB!70!black]
    at (axis cs:0.44,0.22) {$C_B$};

\node[intervention label, rotate=-40, anchor=south]
    at (axis cs:0.18,1.25) {$\Delta\beta=h(p/2)$};
\node[intervention label, rotate=90, anchor=north]
    at (axis cs:0.35,1.40) {$p=1/3$};
\node[intervention label, anchor=south]
    at (axis cs:0.49,0.59) {$\Delta\beta=h(1/3)$};
\node[intervention label, rotate=-20, anchor=south]
    at (axis cs:0.83,0.42) {$\Delta\beta=h(p/2)$};

% Finite-population example: intervention share 0.34 and intensity 2.
\addplot[
    only marks, mark=*, mark size=2.5pt,
    mark options={fill=examplePoint, draw=white, line width=0.7pt}
] coordinates {(0.34,2)};
\node[intervention label, anchor=west]
    at (axis cs:0.38,2) {$(0.34,2)$};

% Both intervention controls increase toward the northeast.
\draw[-{Latex[length=2.1mm, width=1.5mm]}, line width=1.1pt,
    draw=coordinatedA!65!black]
    (axis cs:0.73,1.50) -- (axis cs:0.94,2.17);
\node[
    text=coordinatedA!65!black, font=\scriptsize,
    anchor=south, align=center
] at (axis cs:0.70,2.33)
    {Larger intervention share\\and intervention intensity};

\end{axis}
\end{tikzpicture}

\caption{
Selection by intervention share $p$ and intervention intensity $\Delta\beta$
for the game in Eq.~\eqref{eq:motivating-game}, with initial risk parameter
$\beta_L=0$ (so $\tau_L=2/3$ and $\tau_H=\tau(\Delta\beta)$).
Colored regions indicate the unique SSE for sufficiently large admissible
populations away from the dashed boundaries; $h=\tau^{-1}$.
The marked point $(p,\Delta\beta)=(0.34,2)$ represents raising the risk parameter
of $34$ of $100$ agents from $\beta_L=0$ to $\beta_H=2$.
}
\label{fig:intervention-phase-diagram}
\end{figure}

For the game in Eq.~\eqref{eq:motivating-game}, agents initially have
$\beta_L=0$ and $\tau_L=2/3$. Assigning $\beta_H=2$ to the modified agents
yields an intervention intensity of $\Delta\beta=2$ and a threshold of
\[
\tau_H=\tau(2)=\frac{e^{-4}+e^{-6}}{1+e^{-4}+e^{-6}}\approx0.02037.
\]
Since $\tau_H<1-\tau_L=1/3$, the intervention intensity
condition~\eqref{eq:intervention-intensity-condition} holds. The intervention share
condition~\eqref{eq:intervention-share-condition} then requires $p>1/3$:
more than one third of the existing agents must be assigned $\beta_H=2$
to make $C_A$ the unique SSE in the large-population limit.

For $N=100$, the integer thresholds are $K_H=2$ and $K_L=66$.
The finite-population intervention intensity condition in Corollary~\ref{cor:finite-control},
$K_H+1<\max\{N-K_L,\,n-K_H\}$, holds for every $n$ because $K_H+1=3<34=N-K_L$.
The intervention share condition $n>2K_L+1-N$ in the same corollary therefore gives the minimum number of
agents whose risk attitudes must be modified:
$n^\star=2K_L+2-N=34$.
With intervention share $p=0.34$ and intervention intensity $\Delta\beta=2$,
$C_A$ is the unique SSE.
Fig.~\ref{fig:intervention-phase-diagram} shows the selection regions
for $\beta_L=0$ and marks this example at $(p,\Delta\beta)=(0.34,2)$.
Moving northeast in the figure increases both the intervention share $p$
and intervention intensity $\Delta\beta$.

Adding $M$ new agents at the same intervention intensity $\Delta\beta=2$
(so $\beta_H=2$) gives $N=100+M$, $n=M$, and $K_L=\lfloor2(100+M)/3\rfloor$.
The intervention share condition in Corollary~\ref{cor:finite-control} then
reduces to $M>50$ for integer $M$.
At $M=51$, $K_H=3$ and $K_L=100$, so the intervention intensity
condition also holds because $4<\max\{51,48\}$.
Thus, successful intervention requires adding at least $51$ new agents.
As discussed in Section~\ref{subsec:modifying-versus-adding}, modifying
existing agents' risk attitudes requires the planner to influence fewer agents than adding new
agents to achieve efficient selection at the same intervention intensity.

Finally, at fixed intervention intensity $\Delta\beta=2$, we have
$2\tau_H\approx0.04074$ and $2\tau_L-1=1/3$. As the intervention share $p$
increases, the uniquely selected equilibrium in the large-population limit is $C_B$ for
$p<2\tau_H$, $P$ for $2\tau_H<p<1/3$, and $C_A$ for $p>1/3$.
For these payoffs, Proposition~\ref{prop:welfare-polarized-sse} gives
\[
p_B=\frac{2d-b-c}{a-b-c+d}=\frac13=2\tau_L-1.
\]
Its sufficient condition therefore reduces to $2\tau_H<p<1/3$, covering
the entire open interval where $P$ is uniquely selected. Throughout this
interval, the proposition implies
$W(P)<W(C_B)=2<4=W(C_A)$.
Thus, at this intervention intensity, moving long-run behavior from $C_B$
into the polarized region reduces welfare below the pre-intervention
level; the intervention share must exceed $1/3$ to select $C_A$ and
raise welfare to $4$.
This highlights the importance of the planner's two control variables:
intervention intensity and intervention share. As shown in
Fig.~\ref{fig:intervention-phase-diagram}, selecting $C_A$ as the unique SSE
requires both a sufficiently high intervention intensity and a sufficiently
large intervention share.

\section{Concluding Remarks}
\label{sec:conclusion}

We studied evolutionary learning under BRM in coordination games with two
subpopulations, each with a fixed risk attitude. We showed that the heterogeneity
of risk attitudes in learning can make a polarized equilibrium
uniquely stochastically stable, even though coordination on the payoff-dominant
action strictly improves every agent's expected and ERM-adjusted payoffs.
Polarization can also yield lower welfare than either coordinated outcome.
Starting from a homogeneous population, we characterized the joint roles of
intervention share and intensity in promoting efficient selection. Sufficiently
high intensity and a sufficiently large share ensure unique selection of the
payoff-dominant equilibrium, while intermediate intervention shares can instead
select an inefficient polarized equilibrium.

Our analysis assumes two fixed risk attitudes and the BRM protocol, and focuses 
on long-run equilibrium selection without providing convergence-time guarantees. 
Future work could extend the results to populations with more than two risk
attitudes, including fully heterogeneous populations, and to other revision
protocols. Another direction is to quantify the time needed for interventions
to improve outcomes.

\section{AI Disclosure}
GPT was used to assist with writing, checking the manuscript, and generating TikZ figures. The authors thoroughly reviewed and revised all AI-assisted content and take full responsibility for the manuscript.

\bibliographystyle{IEEEtran}
\bibliography{main-references}

\end{document}